\documentstyle[aps]{revtex}

\begin{document}
\draft
\title{$\beta $ decay and shape isomerism in $^{74}$Kr}
\author{P. Sarriguren, E. Moya de Guerra, A. Escuderos, 
and A.C. Carrizo}
\address{Instituto de Estructura de la Materia,\\
Consejo Superior de Investigaciones Cient\'{\i }ficas, \\
Serrano 123, E-28006 Madrid, Spain}
\date{\today }
\maketitle

\begin{abstract}
We study the properties of $^{74}$Kr, and particularly the 
Gamow Teller strength distribution, using a deformed 
selfconsistent HF+RPA method with Skyrme type interactions. 
Results are presented for two density-dependent effective 
two-body interactions, including the dependence on 
deformation of the HF energy that exhibits two minima at 
close energies and distant deformations, one prolate and 
one oblate. We study the role of deformation, residual 
interaction, pairing and RPA correlations on the Gamow 
Teller strength distribution. Results on moments of inertia 
and gyromagnetic factors, as well as on $E0$ and $M1$ 
transitions are also presented.
\end{abstract}


\section{Introduction}

The study of exotic nuclei, characterized by nonoptimal 
$N/Z$ ratios, has attracted an increasing attention during 
the last decades from both experimental and theoretical 
points of view \cite{interest1,interest2}. Nevertheless, 
it is still true that most of the knowledge we have at 
present about nuclear systems corresponds to the nuclei 
inside the valley of stability, where the balance between 
protons and neutrons make these systems stable against 
particle emission. There is still much to be learned about
systems out of this region of stability. Many complementary 
experimental projects and new facilities involving 
radioactive nuclear beams are being carried out or planned 
at CERN, GANIL, Grenoble-Munich, GSI, Louvain-la-Neuve, 
RIKEN... to extend this limited knowledge.

Information on these nuclei is of prime importance to 
understand different problems in nuclear physics related to 
the way nuclei respond as we vary their masses and isospin 
parameters to extreme values of the $N/Z$ ratio. One can 
also learn about specific components of the N-N force and 
the modification of this force in the nuclear medium. Since 
the isospin dependence of the N-N interaction is largely 
unknown, the structure of single particle states, collective 
modes and behaviour of global nuclear properties is very 
uncertain in nuclei with unusual $N/Z$ ratios. The 
theoretical techniques and parameters are optimized to 
reproduce the properties of known atomic masses and it is 
not clear how accurate is the theoretical extrapolation to 
regions far from stability. The study of nuclei far from 
stability provides then new basic pieces of information on 
nuclear structure.

Knowledge of the decay properties of radioactive nuclei is 
also essential for nuclear astrophysics \cite{astro}, where 
one tries to understand the production of energy and the 
synthesis of elements in stars during stellar events. In 
that respect, reactions with radioactive partners become 
very important, specially during violent stellar events with 
extreme conditions of density and temperature because the 
average time between successive nuclear reactions is much 
shorter than the average decay time of the radioactive 
nucleus. Therefore, these radioactive nuclei do not have 
time to decay before participating in new nuclear reactions. 
In particular, the decay properties of proton rich nuclei 
are fundamental to understand the r-p process (rapid proton 
capture nucleosynthesis) that involves multiple proton
capture reactions on proton rich nuclei. The microscopic 
structure of exotic nuclei near the drip lines is therefore 
a topic of great interest in nuclear structure as well as 
in astrophysics.

The neutron deficient side of the valley of stability is 
particularly interesting because in this region a large 
fraction of the Gamow Teller (GT) strength is accessible 
in $\beta ^{+}$decay \cite{hamasag}. Usually kinematics 
limits $\beta $ decay to a narrow energy window that in 
general misses the main part of the giant GT resonance, but 
this is not the case in proton rich nuclei. Therefore, 
proton rich nuclei provide a good opportunity to test the 
GT strength measured by other less direct means such as 
$(p,n)$ reactions, where the GT strength is extracted by 
extrapolating the empirical proportionality between the 
zero degree cross section and the $\beta $ decay strength 
to low-lying GT states \cite{gaarde}.

All of these features, together with the ongoing 
experimental effort to measure GT strengths in proton rich 
medium mass nuclei \cite{miehe}, make this region the 
focus of this paper.

It is clear that, at some point, extensions of the standard 
approaches will be necessary to cover particular aspects 
that, although not important in stable nuclei, may play a 
role in describing the unstable ones. Issues such as the 
role of the continuum \cite{naza} in weakly bound systems 
or the neutron-proton pairing \cite{nppairing} in 
$N\approx Z$ nuclei should be considered. It is nevertheless 
important to know whether the most powerful tools, designed 
to reproduce the properties of nuclei within the valley of
stability, are still valid when one approaches the drip 
lines. One should know the limits of applicability of the 
existing interactions and methods. In our view this has not 
been sufficiently explored yet, particularly in what 
concerns the GT strength distribution.

With this view in mind we study in this paper the 
$\beta ^{+}$ decay of $^{74}$Kr, which is a medium mass 
proton rich nucleus of recent experimental \cite{gelletly} 
and theoretical interest. This nucleus is also 
representative of a mass region that exhibits a wide 
variety of shapes including large quadrupole deformations 
($\beta _{2}\simeq 0.4$) and shape coexistence. Our method 
consists of a selfconsistent formalism based on a deformed 
Hartree-Fock (HF) mean field obtained with a Skyrme 
interaction including pairing correlations in the BCS 
approximation. We add to this mean field a spin-isospin 
residual interaction with a coupling strength derived
by integrating over the nuclear volume the Landau-Migdal 
force, obtained from the same Skyrme interaction. The 
residual force is therefore consistent with the mean field. 
The equations of motion are solved in the proton-neutron 
quasiparticle random phase approximation (QRPA) \cite{qrpa}.
Most $\beta $ decay calculations in the past used as input 
in one way or another experimental single particle levels, 
but the extrapolation of these empirical information to 
exotic nuclei is at least questionable and the need for 
selfconsistent calculations in unstable nuclei has been 
emphasized (see for instance \cite{nupecc}). There are also 
selfconsistent mean field studies \cite{naza}, as well as 
investigations \cite{hamamoto} of the dependence on 
deformation of the GT strength distributions in Tamm 
Dancoff approximation (TDA), but to our knowledge this is 
the first time that the selfconsistent deformed QRPA 
formalism described here is used to calculate GT strengths. 
A similar formalism for spin $M1$ strength distributions 
has been successfully tested \cite{sar96,sar97} against 
experimental data in the rare earths and actinide nuclei.

The paper is organized as follows: In Section 2 we 
introduce the theoretical framework used to describe the 
$\beta $ decay. We first introduce the selfconsistent 
method for the mean field and residual interactions and 
then we obtain the\ $\beta $ strengths in bare two 
quasiparticle (qp) approximation, in TDA, and in QRPA. 
In Section 3 we study the ground state properties and the 
low-lying structure of $^{74}$Kr and compare our 
calculations to experiment. In Section 4 we discuss the 
results on Gamow Teller strength distributions and sum 
rules obtained in our calculations studying the 
sensitivity of these results to the various ingredients 
of our approach such as the effective two-body interaction, 
the pairing interaction, the nuclear shape, and the RPA 
correlations. We also show results on $M1$ transitions. 
Section 5 contains half-life results in various approaches. 
Section 6 contains the main conclusions of this work.

\section{Brief description of theory and details of 
calculations}

In this section we summarize the theory involved in the 
microscopic calculations presented in Sections 3 and 4. Our 
method is selfconsistent in the sense that, as described in 
Subsection A, with a given two-body density-dependent 
effective interaction we derive: a) the selfconsistent mean 
field generating the single particle energies, wave 
functions and occupations of the ground state, b) the 
particle-hole Landau-Migdal interaction averaged over the 
nuclear volume generating the QRPA modes. Thus, the same 
two-body interaction is used to derive the QRPA excitations 
consistently with the quasiparticle basis and ground state, 
as described in Subsection B.

\subsection{Mean field calculations and residual 
interactions}

It is well known that the density-dependent HF approximation 
gives a very good description of ground-state properties for 
both spherical and deformed nuclei \cite{flocard} and it is 
at present the most reliable mean field description. We 
consider in this paper two different Skyrme forces. On the 
one hand we use the Skyrme force Sk3 \cite{beiner} because 
it is the most extensively used Skyrme force and we consider 
it as a reference. We use Sk3 in its density dependent 
two-body version that has better spin-isospin properties 
than the three-body one \cite{sar96,sar97}. On the other 
hand we use the SG2 force \cite{giai} of Van Giai and 
Sagawa. The two forces were designed to fit ground state 
properties of spherical nuclei and nuclear matter properties 
but, in addition, the force SG2 gives a good description
of Gamow Teller excitations in spherical nuclei \cite{giai}. 
Recently \cite{sar96}, we applied these two forces and a 
similar method to make a rather extensive study of isoscalar 
and isovector spin M1 excitations in deformed nuclei 
obtaining a good description of the available data, 
particularly with SG2. This gives us confidence on the 
predictive power of the method and on the reliability of the 
SG2 force when considering isovector spin properties of 
deformed nuclei. The parameters of these two interactions 
are given in Table 1, and the corresponding HF energy 
density functional for an even-even nucleus has the form 
\begin{eqnarray}
{\cal E}\left( {\bf r}\right) &=&\sum_{st}\rho _{st}
\sum_{s^{\prime }t^{\prime }}\left\{ \frac{1}{2}t_{0}
\rho _{s^{\prime }t^{\prime }}\left[ 1-\delta 
_{ss^{\prime }}\delta _{tt^{\prime }}+x_{0}\left( \delta
_{ss^{\prime }}-\delta _{tt^{\prime }}\right) \right] 
\right.  \nonumber \\
&&+\frac{1}{4}t_{2}\left( \tau _{s^{\prime }t^{\prime }}+
\frac{1}{4}\nabla ^{2}\rho _{s^{\prime }t^{\prime }}\right) 
\left[ 1+\delta _{ss^{\prime }}\delta _{tt^{\prime }}+
x_{2}\left( \delta _{ss^{\prime }}+\delta _{tt^{\prime }}
\right) \right]  \nonumber \\
&&+\frac{1}{16}t_{1}\left( 4\tau _{s^{\prime }t^{\prime }}
-3\nabla ^{2}\rho _{s^{\prime }t^{\prime }}\right) \left[ 
1-\delta _{ss^{\prime }}\delta _{tt^{\prime }}+x_{1}\left( 
\delta _{ss^{\prime }}-\delta _{tt^{\prime }}\right) 
\right]  \nonumber \\
&&+\frac{1}{12}t_{3}\rho ^{\alpha }\rho _{s^{\prime }
t^{\prime }}\left[ 1-\delta _{ss^{\prime }}
\delta _{tt^{\prime }}+x_{3}\left( \delta
_{ss^{\prime }}-\delta _{tt^{\prime }}\right) \right]  
\nonumber \\
&&\left. +\frac{i}{2}W_{0}\bbox{\nabla}{\bf \cdot J}_
{s^{\prime }t^{\prime }}\left( 1+\delta _{tt^{\prime }}
\right) \right\} +{\cal E}_{C}\left( {\bf r}\right)  
\label{dens_ener}
\end{eqnarray}
with ${\cal E}_{C}$ the Coulomb energy density

\begin{equation}
{\cal E}_{C}\left( {\bf r}\right) =e^{2}\frac{1}{2}
\int d{\bf r}^{\prime }\frac{\rho _{p}\left( {\bf r}^
{\prime } \right) \rho _{p}\left( {\bf r}\right) }
{\left| {\bf r}- {\bf r}^{\prime }\right| }-\frac{3}{4}
e^{2}\rho _{p} \left( {\bf r}\right) \left[ \frac{3}{\pi }
\rho _{p} \left( {\bf r}\right) \right] ^{1/3}  
\label{Coulomb}
\end{equation}
The spin-isospin $\left( st\right) $ components of the 
nucleon, kinetic energy, and magnetization densities are

\begin{equation}
\rho _{st}\left( {\bf r}\right) =\sum_{i}v_{i}^{2}\left| 
\phi _{i}\left( {\bf r},s,t\right) \right| ^{2}  
\label{rho_st}
\end{equation}
\begin{equation}
\tau _{st}\left( {\bf r}\right) =\sum_{i}v_{i}^{2}\left| 
\bbox{\nabla}\phi _{i}\left( {\bf r},s,t\right) 
\right| ^{2}  \label{tau_st}
\end{equation}

\begin{equation}
{\bf J}_{st}\left( {\bf r}\right) =\sum_{i,s^{\prime }} 
v_{i}^{2}\phi _{i}^{*}\left( {\bf r},s^{\prime },t\right) 
\left( -i\bbox{\nabla}\times \bbox{\sigma}\right) 
\phi _{i}\left( {\bf r},s,t\right)  \label{J_st}
\end{equation}
with

\begin{equation}
\rho _{t}=\sum_{s}\rho _{st}\; ,  \label{rhot}
\end{equation}

\begin{equation}
\rho =\sum_{t=p,n}\rho _{t}  \label{rhotot}
\end{equation}
and similarly for $\tau $ and ${\bf J}$.

In our calculations time-reversal and axial symmetry are 
assumed. The single-particle wave functions are expanded 
in terms of the eigenstates of an axially symmetric 
harmonic oscillator in cylindrical coordinates. The 
single-particle states $\left| i\right\rangle $ and their 
time reversed $\left| \bar{i}\right\rangle $ are 
characterized by the eigenvalues $\Omega $ of $J_{z}$ , 
parity $\pi _i$, and energy $\epsilon _{i}$

\begin{equation}
\left| i\right\rangle =\sum_{N}\frac{\left( -1\right) ^{N}
+ \pi _{i}}{2}\sum_{n_{r},n_{z},\Lambda \geq 0,\Sigma } 
C_{Nn_{r}n_{z}\Lambda \Sigma }^{i}\left| Nn_{r}n_{z}
\Lambda \Sigma \right\rangle   \label{sp_i}
\end{equation}
with $\Omega _{i}=\Lambda +\Sigma \geq \frac{1}{2}$, and

\begin{equation}
\left| \bar{i}\right\rangle =\sum_{N}\frac{\left( -1 
\right) ^{N}+\pi _{i}}{2} \sum_{n_{r},n_{z},\Lambda 
\geq 0, \Sigma }C_{Nn_{r}n_{z}\Lambda \Sigma }^{i}
\left( -1 \right) ^{\frac{1}{2}-\Sigma }\left| 
Nn_{r}n_{z}-\Lambda -\Sigma \right\rangle  \label{sp_ibar}
\end{equation}
with $\Omega _{\bar{i}}=-\Omega _{i}=-\Lambda -\Sigma 
\leq -\frac{1}{2}$. For each $N$ the sum over 
$n_{r},n_{z},\Lambda \geq 0$ is extended to the
quantum numbers satisfying $2n_{r}+n_{z}+\Lambda =N.$ The 
sum over $N$ goes from $N=0$ to $N=10$ in our calculations.

The single-particle energies $\epsilon _{i}$ and 
wave functions are obtained from the HF equations

\begin{equation}
\frac{\delta {\cal E}}{\delta \phi _{i}^{*}}=
\epsilon _{i} \phi _{i}
\label{hf}
\end{equation}
We use the McMaster code that is based in the formalism 
developed in Ref.\cite{vautherin} as described in 
Ref.\cite{vallieres}. The method includes pairing between 
like nucleons in the BCS approximation with fixed gap
parameters for protons, $\Delta _{p},$ and neutrons, 
$\Delta _{n}$. The number equation in the neutron 
sector reads

\begin{equation}
2\sum_{i}v_{i}^{2}=N  \label{numeq}
\end{equation}
where $v_{i}^{2}$ are the occupation probabilities

\begin{equation}
v_{i}^{2}=\frac{1}{2}\left[ 1-\frac{\epsilon _{i}- 
\lambda _{n}}{E_{i}}\right] \;;\;\;u_{i}^{2}=1-v_{i}^{2}  
\label{occu}
\end{equation}
in terms of the quasiparticle energies

\begin{equation}
\;E_{i}=\sqrt{\left( \epsilon _{i}-\lambda _{n} 
\right) ^{2}+\Delta _{n}^{2}}
\label{qpener}
\end{equation}
Eq. (\ref{numeq}) is solved iteratively at the end of each 
Hartree-Fock iteration to determine the Fermi level 
$\lambda _{n}$. Similar equations are used to determine the 
Fermi level and occupation probabilities for protons by 
changing $N$ into $Z$, $\Delta _{n}$ into $\Delta _{p}$, and 
$\lambda _{n} $ into $\lambda _{p}$.

The fixed gap parameters are determined phenomenologically 
from the odd-even mass differences through a symmetric five 
term formula involving the experimental binding energies 
\cite{audi}:

\begin{eqnarray}
\Delta _{n} &=&\frac{1}{8}\left[ B\left( N-2,Z\right) -4B
\left( N-1,Z\right) +6B\left( N,Z\right) \right.  
\nonumber \\
&&\left. -4B\left( N+1,Z\right) +B\left( N+2,Z\right) 
\right]  \label{gaps}
\end{eqnarray}
A similar expression is found for the proton gap 
$\Delta _{p}$ by changing $N $ by $Z$ and vice versa. 
For $^{74}$Kr we obtain $\Delta _{n}=\Delta _{p}=1.5 $ MeV.

The HF theory gives a single solution which is the Slater 
determinant of lowest energy. To allow for shape coexistence 
one has to extend the theory to a constrained HF theory 
\cite{constraint}. Minimization of the HF energy under the 
constraint of holding the nuclear deformation fixed is 
carried out over a range of deformations. When more than 
one local minimum occurs for the total energy as a 
function of deformation, shape coexistence results. The 
energy surfaces as a function of deformation that we will 
discuss in Section 3, are obtained by this procedure 
including a quadratic quadrupole constraint 
\cite{constraint}.

Following Bertsch and Tsai \cite{bertsch} the particle-hole 
interaction consistent with the HF mean field can be 
obtained as

\begin{equation}
V_{ph}=\frac{1}{16}\sum_{sts^{\prime }t^{\prime }}\left[ 
1+\left( -1\right) ^{s-s^{\prime }}\bbox{\sigma_1 \cdot 
\sigma_2}\right] \left[ 1+\left( -1\right) ^{t-t^{\prime }}
\bbox{\tau_1 \cdot \tau_2}\right] \frac{\delta ^{2}
{\cal E}}{\delta \rho _{st}\left( {\bf r}_{1}\right) \delta 
\rho _{s^{\prime }t^{\prime }}\left( {\bf r}_{2}\right) }  
\label{Vph}
\end{equation}
This gives a local interaction that can be put in the 
Landau-Migdal form \cite{migdal}. For the study of $\beta $ 
decay the relevant residual interactions are the isospin 
contact forces generating the allowed Fermi transitions 
$(\Delta L=0,\Delta S=0,\Delta I^{\pi }=0^{+})$

\begin{equation}
V_{F}\left( 12\right) =\chi _{F}\;\left(
t_{1}^{+}t_{2}^{-}+t_{1}^{-}t_{2}^{+}\right)  \label{V_F}
\end{equation}
and the spin-isospin contact forces generating the allowed 
Gamow Teller transitions 
$(\Delta L=0,\Delta S=1,\Delta I^{\pi }=1^{+})$

\begin{equation}
V_{GT}\left( 12\right) =\chi _{GT}\;\bbox{\sigma_1 \cdot 
\sigma_2}\left( t_{1}^{+}t_{2}^{-}+t_{1}^{-}t_{2}^{+}
\right)  \label{V_G}
\end{equation}
where we use the convention $t^{+}\left| p\right\rangle =
\left| n\right\rangle ,\; t^{-}\left| n\right\rangle =
\left| p\right\rangle .$ The latter $\left( V_{GT}\right) $ 
is the charge changing component of the spin-spin 
interaction $H_{SS}$

\begin{equation}
H_{SS}=\frac{K_{S}}{4A}\left[ \left( 1+q\right) {\bf s}_{1} 
{\bf \cdot s}_{2}+\left( 1-q\right) {\bf s}_{1} {\bf 
\cdot s}_{2}\;\bbox{\tau_1 \cdot \tau_2}\right]  \label{hss}
\end{equation}
considered in Refs. \cite{sar96,sar97} for the study of spin 
$M1$ excitations. Not allowed transitions 
$\left( \Delta L>0\right) $ produce strengths which are 
orders of magnitude smaller than the allowed ones 
$\left( \Delta L=0\right) $ and will not be considered in 
this work.

After functional differentiation we get from Eqs. 
(\ref{dens_ener}) and (\ref{Vph}-\ref{V_G}), assuming 
symmetric uniform nuclear matter and averaging over the 
nuclear volume $V$

\begin{equation}
\chi _{F}=\frac{3}{4\pi R^{3}}\left( -\frac{1}{2}\right) 
\left\{ t_{0}\left( 1+2x_{0}\right) -\frac{1}{2}k_{F}^{2} 
\left[ t_{2}\left( 1+2x_{2}\right) -t_{1}\left( 1+2x_{1} 
\right) \right] +\frac{1}{6}t_{3}\rho ^{\alpha }\left(
1+2x_{3}\right) \right\}  \label{X_F}
\end{equation}

\begin{equation}
\chi _{GT}=\frac{3}{4\pi R^{3}}\left( -\frac{1}{2}\right) 
\left\{ t_{0}+\frac{1}{2}k_{F}^{2}\left( t_{1}-t_{2}\right) 
+\frac{1}{6}t_{3}\rho ^{\alpha }\right\}  \label{X_GT}
\end{equation}
where $R$ is the nuclear radius and $k_{F}$ the Fermi 
momentum $k_{F}=\left( 3\pi ^{2}\rho /2\right) ^{1/3}$. 
These coupling strengths are related to the familiar 
Landau-Migdal parameters $F_{0\text{ }}^{\prime } $ 
and $G_{0\text{ }}^{\prime }$ (see for instance 
\cite{giai}) by

\begin{equation}
\chi _{F}=\frac{2F_{0}^{\prime }}{VN_{0}}\;,\;\chi _{GT}= 
\frac{2G_{0}^{\prime }}{VN_{0}}  \label{XFG}
\end{equation}
where $V=4\pi R^{3}/3$ and $N_{0}=\left( 2m^{\star }
k_{F}/\hbar ^{2}\pi ^{2}\right) $, with $m^{\star }$ the 
effective mass. For completeness the values of $\chi _{F}$ 
and $\chi _{GT}$ are also given in Table 1.

\subsection{QRPA equations and $\beta $ decay strengths}

Let us first consider Gamow Teller excitations. Using the 
notation $\beta _{K}^{\pm }=\sigma _{K}t^{\pm }\; 
\left( K=0,\pm 1\right) $, we write Eq. (\ref{V_G}) in the 
more traditional form

\begin{equation}
V_{GT}=2\chi _{GT}\sum_{K}\left( -1\right) ^{K}\; 
\beta _{K}^{+}\beta _{-K}^{-}  \label{betbet}
\end{equation}
where in second quantization we can write

\begin{eqnarray}
\beta _{K}^{+} &=&\sum_{np}\left\langle n\left| 
\sigma _{K}\right| p\right\rangle a_{n}^{+}a_{p}^{+}=
\sum_{np}\left\langle n\left| \sigma _{K}\right| p
\right\rangle \left\{ u_{n}v_{p}\alpha _{n}^{+}
\alpha _{\bar{p}}^{+}\right.  \nonumber \\
&&\left. +v_{n}u_{p}\alpha _{\bar{n}}\alpha _{p}+
u_{n}u_{p}\alpha _{n}^{+}\alpha _{p}+v_{n}v_{p}
\alpha _{\bar{p}}^{+}\alpha _{\bar{n}}\right\}
\label{betak}
\end{eqnarray}
and $\beta _{-K}^{-}=\left( -1\right) ^{K}\left( 
\beta _{K}^{+}\right) ^{+}$. We use the standard notation 
$a^{+}\left( a\right) $ for particle creation (annihilation) 
and $\alpha ^{+}\left( \alpha \right) $ for quasiparticle
creation (annihilation) operators. We write the 
proton-neutron QRPA phonon operator for Gamow Teller 
excitations in even-even nuclei

\begin{equation}
\Gamma _{\omega _{K}}^{+}=\sum_{\gamma _{K}}\left[ 
X_{\gamma _{K}}^{\omega _{K}}A_{\gamma _{K}}^{+}-
Y_ {\gamma _{K}}^{\omega _{K}}A_{\bar{\gamma}_{K}}
\right]  \label{phon}
\end{equation}
where $A_{\gamma _{K}}^{+}\;,\;A_{\bar{\gamma}_{K}}$ are 
two-quasiparticle operators of the form

\begin{equation}
A_{\gamma _{K}}^{+}=\alpha _{n}^{+}\alpha _{\bar{p}}^{+}  
\label{aplus}
\end{equation}

\begin{equation}
A_{\bar{\gamma}_{K}}=\left( -\alpha _{\bar{n}}^{+} 
\alpha _{p}^{+}\right) ^{+}=\alpha _{\bar{n}}
\alpha _{p}  \label{abar}
\end{equation}
with $\Omega _{n}+\Omega _{\bar{p}}=K$, where 
$n\left( p\right) $ stands for any neutron (proton) states 
$\left| i\right\rangle $ or $\left| \bar{i}\right\rangle $ 
and $\bar{n}\left( \bar{p}\right) $ is its time reverse. 
The QRPA equations for the even-even system are

\begin{equation}
\left\langle \phi _{0}\left| A_{\gamma _{K}}\left[ H, 
\Gamma _{\omega _{K}}^{+}\right] \right| \phi _{0} 
\right\rangle =\omega _{K}\left\langle \phi _{0}\left| 
A_{\gamma _{K}}\Gamma _{\omega _{K}}^{+}\right| \phi _{0}
\right\rangle  \label{rpa1}
\end{equation}

\begin{equation}
\left\langle \phi _{0}\left| \left[ H, \Gamma 
_{\omega _{K}}^{+}\right] A_{\bar{\gamma}_{K}}^{+} 
\right| \phi _{0}\right\rangle =\omega _{K}\left\langle 
\phi _{0}\left| \Gamma _{\omega _{K}}^{+} 
A_{\bar{\gamma}_{K}}^{+}\right| \phi _{0}\right\rangle  
\label{rpa2}
\end{equation}
with

\begin{equation}
H=\sum_{n}\alpha _{n}^{+}\alpha _{n}E_{n}+\sum_{p}
\alpha _{p}^{+}\alpha _{p}E_{p}+V_{GT}-\left\langle 
\phi _{0}\left| H\right| \phi _{0}\right\rangle
\label{h_rpa}
\end{equation}
$\left| \phi _{0}\right\rangle $ is the HF+BCS ground state 
satisfying $\alpha _{t}\left| \phi _{0}\right\rangle =0$ 
for any $t$. From these equations the forward and backward 
amplitudes are obtained as

\begin{equation}
X_{\gamma _{K}}^{\omega _{K}}=\frac{2\chi _{GT}} 
{\omega _{K}-{\cal E}_{\gamma _{K}}}\left( a_{\gamma _{K}} 
{\cal M}_{+}^{\omega _{K}}+b_{\gamma _{K}}{\cal M}_{-}^ 
{\omega _{K}}\right)  \label{xwa}
\end{equation}

\begin{equation}
Y_{\gamma _{K}}^{\omega _{K}}=\frac{-2\chi _{GT}} 
{\omega _{K}+{\cal E}_{\gamma _{K}}}\left( b_{\gamma _{K}} 
{\cal M}_{+}^{\omega _{K}}+a_{\gamma _{K}} 
{\cal M}_{-}^{\omega _{K}}\right)  \label{ywa}
\end{equation}
with ${\cal E}_{\gamma _{K}}=E_{n}+E_{p}$ the 
two-quasiparticle excitation energy with $E_{i}$ given by 
Eq. (\ref{qpener}), ${\cal M}_{\pm }^{\omega _{K}}$ are

\begin{equation}
{\cal M}_{+}^{\omega _{K}}=\sum_{\gamma _{K}}\left( 
a_{\gamma _{K}}X_{\gamma _{K}}^{\omega _{K}}+b_{\gamma _{K}} 
Y_{\gamma _{K}}^{\omega _{K}}\right) \label{mplus}
\end{equation}

\begin{equation}
{\cal M}_{-}^{\omega _{K}}=\sum_{\gamma _{K}}\left( 
b_{\gamma _{K}}X_{\gamma _{K}}^{\omega _{K}}+a_{\gamma _{K}} 
Y_{\gamma _{K}}^{\omega _{K}}\right) \label{mminus}
\end{equation}
with

\begin{equation}
a_{\gamma _{K}}=u_{n}v_{p}\Sigma _{K}^{np}  \label{aalf}
\end{equation}

\begin{equation}
b_{\gamma _{K}}=v_{n}u_{p}\Sigma _{K}^{np}  \label{balf}
\end{equation}
and

\begin{equation}
\Sigma _{K}^{np}=\left\langle n\left| \sigma _{K}\right| 
p\right\rangle \label{sigma}
\end{equation}

Owing to the transformation properties under time reversal 
and hermitian conjugation it is easy to show that: i) the 
$K=-1$ and $K=1$ modes are related to each other through 
time reversal and are degenerate; ii) the $K=0$ modes are 
time odd and satisfy $X_{\bar{\gamma}_{0}}^{\omega _{0}}
=-X_{\gamma _{0}}^{\omega _{0}}$ and 
$Y_{\bar{\gamma}_{0}}^ {\omega _{0}}=
-Y_{\gamma _{0}}^{\omega _{0}}$. Forward and backward 
amplitudes are real. This allows to simplify the 
calculations to include only neutron-proton pairs with 
$\Omega _{n}>0$, i.e., the complete set of $\gamma _{K}$ 
neutron proton pairs can be reduced to the basic sets of 
pairs:

\begin{itemize}
\item  $i_{0}=n_{i}p_{i}$ with $\Omega _{n_{i}}=
\Omega _{p_{i}}\geq \frac{1}{2}$

\item  $i_{-1}=n_{i}p_{i}$ with $\Omega _{n_{i}}=
\Omega _{p_{i}}-1\geq \frac{1}{2}$

\item  $i_{+1}=\left\{ 
\begin{array}{l}
n_{i}p_{i}\;\text{with }\Omega _{n_{i}}=\Omega _{p_{i}}+
1\geq \frac{3}{2} \\ 
n_{i}\bar{p}_{i}\;\text{with }\Omega _{n_{i}}=
\Omega _{p_{i}}=\frac{1}{2}
\end{array}
\right. $
\end{itemize}

The matrix elements are

\begin{equation}
\Sigma _{K}^{n_{i}p_{i}}=\sum_{Nn_{z}\Lambda \Sigma } 
C_{Nn_{z}\Lambda \Sigma +K}^{n_{i}}C_{Nn_{z}\Lambda 
\Sigma }^{p_{i}}\left( 2\Sigma \right) \sqrt{1+\left| 
K\right| }  \label{sig1}
\end{equation}

\begin{equation}
\Sigma _{K=1}^{n_{i}\bar{p}_{i}}=\sum_{Nn_{z}}C_{Nn_{z}0 
\frac{1}{2}}^{n_{i}}C_{Nn_{z}0\frac{1}{2}}^{p_{i}}
\left( -\sqrt{2}\right)  \label{sig2}
\end{equation}
We therefore obtain the following dispersion relations for 
the $K=0$ and $K=1 $ modes:

\begin{eqnarray}
\left( \frac{1}{4\chi _{GT}}\right) ^{2} &=&\frac{1}{2 
\chi _{GT}}\sum_{i_{0}}\frac{\left( a_{i_{0}}^{2}+ 
b_{i_{0}}^{2}\right) }{\omega _{0}^{2}- {\cal E}_
{i_{0}}^{2}}{\cal E}_{i_{0}}+\left( \sum_{i_{0}}a_{i_{0}}
b_{i_{0}}\frac{2{\cal E}_{i_{0}}} {\omega _{0}^{2}-
{\cal E}_{i_{0}}^{2}}\right) ^{2}  \nonumber \\
&&-\sum_{i_{0}}\left( \frac{a_{i_{0}}^{2}}{\omega _{0}+ 
{\cal E}_{i_{0}}}-\frac{b_{i_{0}}^{2}}{\omega _{0}- 
{\cal E}_{i_{0}}}\right) \sum_{i_{0}}\left( 
\frac{b_{i_{0}}^{2}}{\omega _{0}+{\cal E}_{i_{0}}}- 
\frac{a_{i_{0}}^{2}}{\omega _{0}-{\cal E}_{i_{0}}}\right)  
\label{k0}
\end{eqnarray}
for $K=0$, and

\begin{eqnarray}
\left( \frac{1}{2\chi _{GT}}\right) ^{2} &=&\frac{1} 
{\chi _{GT}}\sum_{i_{\rho }\left( \rho =\pm 1\right) } 
\frac{\left( a_{i_{\rho }}^{2}+b_{i_{\rho }}^{2}\right) } 
{\omega _{1}^{2}-{\cal E}_{i_{\rho }}^{2}} 
{\cal E}_{i_{\rho }}+\left( \sum_{i_{\rho }\left( 
\rho =\pm 1\right) }a_{i_{\rho }}b_{i_{\rho }} 
\frac{2{\cal E}_{i_{\rho }}}{\omega _{1}^{2}- 
{\cal E}_{i_{\rho }}^{2}}\right) ^{2}  \nonumber \\
&&-\sum_{i_{\rho }\left( \rho =\pm 1\right) }\left( 
\frac{a_{i_{\rho }}^{2}}{\omega _{1}+{\cal E}_{i_{\rho }}} 
-\frac{b_{i_{\rho }}^{2}}{\omega _{1}-{\cal E}_{i_{\rho }}} 
\right) \sum_{i_{\rho }\left( \rho =\pm 1\right) }\left( 
\frac{b_{i_{\rho }}^{2}}{\omega _{1}+{\cal E}_{i_{\rho }}} 
-\frac{a_{i_{\rho }}^{2}}{\omega _{1}-{\cal E}_{i_{\rho }}} 
\right)  \label{k1}
\end{eqnarray}
for $K=1$.

The forward and backward amplitudes are then obtained from 
Eqs. (\ref{xwa}) and (\ref{ywa}) with the following 
normalization conditions

\begin{equation}
2\sum_{i_{0}}\left[ \left( X_{i_{0}}^{\omega _{0}} 
\right) ^{2}-\left( Y_{i_{0}}^{\omega _{0}}\right) ^{2} 
\right] =1  \label{norm0}
\end{equation}

\begin{equation}
\sum_{i_{\rho }\left( \rho =\pm 1\right) } \left( 
X_{i_{\rho }}^{\omega _{1}}\right) ^{2}- \left( 
Y_{i_{\rho }}^{\omega _{1}}\right) ^{2}=1
\label{norm1}
\end{equation}
Using the inverse transformation

\begin{equation}
A_{i_{K}}^{+}=\sum_{\omega _{K}} \left( X_{i_{K}}^
{\omega _{K}}\Gamma _{\omega _{K}}^{+}+ Y_{i_{K}}^
{\omega _{K}}\Gamma _{\bar{\omega}_{K}}\right)
,  \label{areverse}
\end{equation}
where $\Gamma _{\bar{\omega}_{K}}$ stands for the time 
reverse of $\Gamma _{\omega _{K}},$ and the RPA 
conditions $\Gamma _{\omega _{K}}\left| 0\right\rangle 
=0,$ $\left| \omega _{K}\right\rangle =\Gamma _{\omega
_{K}}^{+} \left| 0\right\rangle ,$ with $\left| 
0\right\rangle $ the QRPA correlated ground state, one 
gets the $\beta _{K}^{\pm }$ strengths

\begin{equation}
\left\langle \omega _{K}\left| \beta _{K}^{\pm }\right| 0 
\right\rangle ={\cal M}_{\pm }^{\omega _{K}}  
\label{strength}
\end{equation}
with

\begin{equation}
{\cal M}_{+}^{\omega _{0}}=2\sum_{i_{0}}\left( a_{i_{0}} 
X_{i_{0}}^{\omega _{0}}+b_{i_{0}}Y_{i_{0}}^{\omega _{0}} 
\right)  \label{mpl0}
\end{equation}

\begin{equation}
{\cal M}_{+}^{\omega _{1}}=\sum_{i_{\rho }\left( \rho =
\pm 1\right) }\left( a_{i_{\rho }} X_{i_{\rho }}^{\omega 
_{1}}+b_{i_{\rho }} Y_{i_{\rho }}^{\omega _{1}}\right)  
\label{mpl1}
\end{equation}
${\cal M}_{-}^{\omega _{K}}$ is obtained from 
${\cal M}_{+}^{\omega _{K}}$ exchanging $a_{i_{K}}$ and 
$b_{i_{K}}$ (see Eq. (\ref{mminus})).

To finish this section we note that the Fermi transitions, 
although not shown here, can be handled in a similar way 
to the $K=0$ GT transitions with the important difference 
that the phonon operator for the Fermi mode is time even. 
Therefore, if we denote by $\omega _{F}$ the frequencies 
of the Fermi excitations one has that 
$X_{\bar{i}_{0}}^{\omega _{F}}=X_{i_{0}}^{\omega _{F}},\; 
Y_{\bar{i}_{0}}^{\omega _{F}}=Y_{i_{0}}^{\omega _{F}}$ and 
we may consider only pairs of the type $i_{0}$ with the 
matrix elements $a_{i_{0}},\;b_{i_{0}}$ replaced by

\begin{equation}
a_{i_{0}}^{\prime }=\left\langle n_{i}|p_{i}\right\rangle
\;u_{n_{i}}v_{p_{i}}  \label{aprime}
\end{equation}

\begin{equation}
b_{i_{0}}^{\prime }=\left\langle n_{i}|p_{i}\right\rangle
\;v_{n_{i}}u_{p_{i}}  \label{bprime}
\end{equation}
With these replacements the dispersion relation in Eq. 
(\ref{k0}) can also be used to find the frequencies 
$\omega _{F}$ of the Fermi modes and the Fermi strengths 
can be obtained similarly to the $K=0$ GT strengths

\begin{equation}
\left\langle \omega _{F}\left| t^{+}\right| 0\right\rangle 
={\cal M}_{+}^{\omega _{F}}=2\sum_{i_{0}}\left( a_{i_{0}}^
{\prime }X_{i_{0}}^{\omega _{F}}+ b_{i_{0}}^{\prime }
Y_{i_{0}}^{\omega _{F}}\right)  \label{wftp}
\end{equation}

\begin{equation}
\left\langle \omega _{F}\left| t^{-}\right| 0 \right\rangle 
={\cal M}_{-}^{\omega _{F}}=2\sum_{i_{0}} \left( a_{i_{0}}^
{\prime }Y_{i_{0}}^{\omega _{F}}+ b_{i_{0}}^{\prime }
X_{i_{0}}^{\omega _{F}}\right)  \label{wftm}
\end{equation}
with

\begin{equation}
X_{i_{0}}^{\omega _{F}}=\left( a_{i_{0}}^{\prime } 
{\cal M}_{+}^{\omega _{F}}+b_{i_{0}}^{\prime } 
{\cal M}_{-}^{\omega _{F}}\right) \frac{2\chi _{F}} 
{\omega _{F}-{\cal E}_{i_{0}}}  \label{xwf}
\end{equation}

\begin{equation}
Y_{i_{0}}^{\omega _{F}}=-\left( b_{i_{0}}^{\prime } 
{\cal M}_{+}^{\omega _{F}}+a_{i_{0}}^{\prime } 
{\cal M}_{-}^{\omega _{F}}\right) \frac{2\chi _{F}} 
{\omega _{F}+{\cal E}_{i_{0}}}  \label{ywf}
\end{equation}
and

\[
2\sum_{i_{0}}\left( \left| X_{i_{0}}^{\omega _{F}} 
\right| ^{2}-\left| Y_{i_{0}}^{\omega _{F}} 
\right| ^{2}\right) =1 
\]

In Section 4 we show also for comparison the Gamow Teller 
strength distributions obtained in the TDA approximation, 
as well as the uncorrelated two-quasiparticle neutron 
proton GT strengths.

The equations for TDA are easily obtained from the above 
RPA equations by taking everywhere the limit 
$1/\left( \omega _{K}+{\cal E}_{i_{K}}\right)
\rightarrow 0,$ that in particular kills the $Y$ amplitudes 
$\left( Y_{i_{K}}^{\omega _{K}}=0\right) $ and simplifies 
the eigenvalue equations. The $\beta _{K}^{\pm }$ strengths 
are then given by

\begin{equation}
\left\langle \omega _{K}^{TDA}\left| \beta _{K}^{\pm } 
\right| \phi _{0}\right\rangle ={\cal M}_{\pm }^ 
{\omega _{K}^{TDA}}  \label{wtda}
\end{equation}
with

\begin{equation}
{\cal M}_{+}^{\omega _{0}^{TDA}}=2\sum_{i_{0}}a_{i_{0}} 
X_{i_{0}}^{\omega _{0}^{TDA}}  \label{mptda}
\end{equation}

\begin{equation}
{\cal M}_{-}^{\omega _{0}^{TDA}}=2\sum_{i_{0}}b_{i_{0}} 
X_{i_{0}}^{\omega _{0}^{TDA}}  \label{mmtda}
\end{equation}

\[
\;2\sum_{i_{0}}\left( X_{i_{0}}^{\omega _{0}^{TDA}} 
\right) ^{2}=1 
\]
and corresponding relations for $K=1.$

Finally, the uncorrelated two-quasiparticle neutron-proton 
Gamow Teller (Fermi) excitations are obtained in the limit 
$\chi _{GT}=0\;\left( \chi _{F}=0\right) $. Obviously in 
this case the excitation energies are the bare
two-quasiparticle energies $\omega _{K}^{2qp}=
{\cal E}_{\gamma _{K}}=E_{n}+E_{p},$ and the $\beta ^{\pm }$ 
strengths are

\begin{equation}
\left\langle \omega _{K}^{2qp}\left| \beta _{K}^{+} \right| 
\phi _{0}\right\rangle =a_{\gamma _{K}}  \label{bpunc}
\end{equation}

\begin{equation}
\left\langle \omega _{K}^{2qp}\left| \beta _{K}^{-}\right| 
\phi _{0}\right\rangle =b_{\gamma _{K}}  \label{bmunc}
\end{equation}
Similar relations hold for Fermi transitions with $K=0$ and 
$a_{\gamma _{0}},\;b_{\gamma _{0}}$ replaced by 
$a_{\gamma _{0}}^{\prime },\;b_{\gamma _{0}}^{\prime }$ 
(\ref{aprime}) and (\ref{bprime}).

In the uncorrelated case it is straightforward to show that 
the Ikeda sum rule is fulfilled. For Fermi transitions 
one has

\begin{eqnarray}
S_{F}^{-}-S_{F}^{+} &=&\sum_{\omega _{F}}\left| 
\left\langle \omega _{F}\left| t^{-}\right| \phi _{0} 
\right\rangle \right| ^{2}-\sum_{\omega _{F}}\left| 
\left\langle \omega _{F}\left| t^{+}\right| \phi _{0}
\right\rangle \right| ^{2}=\sum_{\gamma _{0}}\left[ 
\left( b_{\gamma _{0}}^{\prime }\right) ^{2}-\left( 
a_{\gamma _{0}}^{\prime }\right) ^{2}\right]  \nonumber \\
&=&2\sum_{i_{0}}\left[ \left( b_{i_{0}}^{\prime } \right) 
^{2}-\left( a_{i_{0}}^{\prime }\right) ^{2} \right] =2
\sum_{n_{i},p_{i}}\left| \left\langle n_{i}| p_{i}
\right\rangle \right| ^{2}\left( v_{n_{i}}^{2}-
v_{p_{i}}^{2}\right) =N-Z  \label{ferik}
\end{eqnarray}
For Gamow Teller transitions one has

\begin{equation}
S_{GT}^{-}-S_{GT}^{+}=\sum_{K=0,\pm 1}\left(
S_{GT,K}^{-}-S_{GT,K}^{+}\right) =6\left[
\sum_{n_{i}}v_{n_{i}}^{2}-\sum_{p_{i}}v_{p_{i}}^{2}\right] 
=3\left( N-Z\right)  \label{gamik}
\end{equation}
where we have used Eqs. (\ref{bpunc}) and (\ref{bmunc}) 
to write

\begin{equation}
S_{GT,K}^{-}-S_{GT,K}^{+}=\sum_{\omega _{K}}\left[ \left| 
\left\langle \omega _{K}\left| \beta ^{-}\right| 
\phi _{0} \right\rangle \right| ^{2}-\left| \left\langle 
\omega _{K} \left| \beta ^{+}\right| \phi _{0}\right\rangle 
\right| ^{2}\right] =\sum_{\gamma _{K}}\left[ \left(
b_{\gamma _{K}}\right) ^{2}-\left( a_{\gamma _{K}} 
\right) ^{2}\right]
\label{gamik2}
\end{equation}

It is a simple matter to show that the Ikeda sum rules hold 
also in the TDA and RPA approximations provided all the 
$\omega $ eigenvalues contained in the basis space are 
included in the sum so that the orthonormalization
conditions

\[
\sum_{\omega _{K}}\left( X_{\gamma _{K}}^{\omega _{K}} 
X_{\gamma _{K}^{\prime }}^{\omega _{K}}- Y_{\gamma _{K}}^
{\omega _{K}}Y_{\gamma _{K}^ {\prime }}^{\omega _{K}}
\right) =\delta _{\gamma _{K}, \gamma _{K}^{\prime
}}\;\;\;\text{in RPA} 
\]
and

\[
\sum_{\omega _{K}^{TDA}}X_{\gamma _{K}}^{\omega _{K}^{TDA}} 
X_{\gamma _{K}^{\prime }}^{\omega _{K}^{TDA}}=\delta _ 
{\gamma _{K},\gamma _{K}^{\prime }}\;\;\;\text{in TDA} 
\]
are satisfied. In practice, the strength functions are 
calculated up to an energy cut $\omega \leq E_{cut}$ such 
that Ikeda's sum rule is satisfied up to a few per thousand.

\section{Ground state properties and low lying structure}

The constrained HF method allows one to get a solution for 
each value of the mass quadrupole $Q_{2}$. In Fig. 1 we 
show the HF energy as a function of deformation for the 
two interactions SG2 and Sk3. The best HF solution at each 
$Q_{2}$ value can be obtained by varying the size and 
deformation parameters \cite{vautherin} of the deformed 
harmonic oscillator basis. As it is seen in Fig. 1 there 
are two minima, one in the prolate sector and one in the 
oblate sector, rather close in energy. Thus, as expected, 
shape coexistence takes place in this nucleus. This is an 
interesting feature that was previously predicted (see for 
instance Refs. \cite{interest1,wood} and references therein) 
and that has been experimentally confirmed \cite{gelletly}. 
It should be noted that Fig. 1 leaves open the question as 
to whether the ground state is oblate or prolate. Sk3 
favors prolate shape for the ground state and oblate for 
the shape isomer, in accordance with previous predictions 
\cite{prola}. On the contrary SG2 force favors an oblate 
ground state and a prolate shape isomer. The results shown 
in Fig. 1 exemplify the situation met when several other 
Skyrme type interactions are used. In particular, the 
deformation energy curve obtained from calculations
performed with the Skyrme force Ska \cite{koehler} shows a 
profile quite similar to that of SG2 with an absolute 
oblate minimum with $Q_{2}\simeq -200 $ fm$^{2}$ and a 
prolate isomer at about 1 MeV with $Q_{2}\simeq 700$ fm 
$^{2} $. On the other hand the force Skm$^{\star }$ 
\cite{bartel} produces a profile analogous to Sk3 with an 
absolute prolate minimum at about $Q_{2}=700 $ fm$^{2},$ 
and an oblate isomer at $Q_{2}=-400$ fm$^{2}$. It is also 
worth mentioning that while the prolate solution appears 
at about the same deformation for all the forces 
considered, the oblate solution is spread from 
$Q_{2}=-200$ fm$^{2}$ up to $Q_{2}=-500$ fm$^{2}$ 
depending on the force considered.

The experimental $\beta _{2}$ values deduced from measured 
$B(E2)$ strengths \cite{heese} vary from $\left| \beta _{2}
\right| =0.24$ to $\left| \beta _{2}\right| =0.40$. These 
are consistent with our calculated parameter of deformation 
$\beta _{2}$ defined in terms of the quadrupole moment 
$Q_{2}$ and the r.m.s. radius $\left\langle r^{2}
\right\rangle $

\begin{equation}
\beta _{2}=\sqrt{\frac{\pi }{5}}\frac{Q_{2}}{A\left\langle
r^{2}\right\rangle }\;.  \label{betadefi}
\end{equation}
The results obtained for $\beta _{2}$ can be seen in Table 
2.

Although in Ref. \cite{gelletly} it is stated that the 
ground state is prolate and the shape isomer is oblate, we 
find that experimental evidence supporting this assertion 
is still lacking and that it would be interesting to 
investigate further this question from the experimental 
side. Of course experimental evidence would require 
measuring the static quadrupole moments of the lowest 
$2^{+}$ states, however some hints can be obtained from a
comparison to data of the theoretical low lying spectra 
and $E0$ transition rates obtained with these two 
interactions. To this end we compare in Fig. 2 the 
experimental \cite{tabor} level spectra up to 12 MeV with 
those obtained for each interaction neglecting shape 
mixing. Clearly, the low lying spectra in Fig. 2 obtained 
with the prolate Sk3 compare better with experiment. We
follow the approximate variation after projection method 
\cite{moya} and write the energy of the $K=0^{+}$ band 
head $\left( J=0^{+}\right) $ as

\begin{equation}
E_{0^{+}}=E_{HF}-\frac{\left\langle J^{2}\right\rangle } 
{2{\cal I}_{cr}}
\label{E0+}
\end{equation}
and the energies of the higher $\left( I=2,4,6,...\right) $ 
members of the rotational band are calculated to lowest 
order in angular momentum \cite
{bohr}

\begin{equation}
E_{I^{+}}-E_{0^{+}}=\frac{I\left( I+1\right) } 
{2{\cal I}_{cr}}  \label{EI+}
\end{equation}
where ${\cal I}_{cr}$ is the cranking moment of inertia,

\begin{equation}
{\cal I}_{cr}=2\sum_{\alpha _{t}\beta _{t}}\frac{\left| 
\left\langle \alpha _{t}\beta _{t}\left| J_{x}\right| 
\phi _{0}\right\rangle \right| ^{2}}{E_{\alpha }+
E_{\beta }}  \label{Icr}
\end{equation}
with $\left| \alpha _{t}\beta _{t}\right\rangle =
\alpha _{\alpha _{t}}^{+}\alpha _{\beta _{t}}^{+}\left| 
\phi _{0}\right\rangle $ any two quasiparticle state 
($t=$ proton, neutron). The values of the Fermi level
for neutrons $\lambda _{n}$ and protons $\lambda _{p}$, 
the values of $\left\langle J^{2}\right\rangle $ and 
${\cal I}_{cr}$, together with the charge r.m.s. radii, 
deformations $\beta _{2}$, HF energies 
$\left( E_{HF}\right) ,$ and $0^{+}$ energies ($E_{0^{+}},$
from Eq. (\ref{E0+})), are given in Table 2 for the two 
minima, oblate and prolate, with the two interactions, SG2 
and Sk3. The experimental \cite{audi} binding energy is 
$B(^{74}$Kr$)=-631.281$ MeV.

In any case, since the two minima are close in energy and 
the barriers are not so high, substantial shape mixing can 
be expected. In Fig. 3 we show that the experimental 
\cite{gelletly} $\rho ^{2}\left( E0\right) $ value is
compatible with both Sk3 and SG2 predictions when shape 
mixing is taken into account. Considering that the 
$\left| 0_{1}^{+}\right\rangle $ and $\left| 0_{2}^{+}
\right\rangle $ states mix to form the observed $0^{+}$ 
ground state $\left| 0_{I}^{+}\right\rangle $ and first 
$0^{+}$ excited state $\left| 0_{II}^{+}\right\rangle $

\begin{equation}
\left| 0_{I}^{+}\right\rangle =\sqrt{\lambda } \left| 
0_{1}^{+}\right\rangle +\sqrt{1-\lambda } \left| 
0_{2}^{+}\right\rangle  \label{0I+}
\end{equation}

\begin{equation}
\left| 0_{II}^{+}\right\rangle =\sqrt{1-\lambda }\left|
0_{1}^{+}\right\rangle -\sqrt{\lambda }\left| 0_{2}^{+} 
\right\rangle  \label{0II+}
\end{equation}
The amplitude of the monopole transition is given by 
\cite{bohr}

\begin{eqnarray}
\rho \left( E0,0_{II}^{+}\rightarrow 0_{I}^{+}\right) 
&=&\frac{1}{eR^{2}}\left\langle 0_{II}^{+}\left| 
\hat{E}0\right| 0_{I}^{+}\right\rangle 
\nonumber \\
&=&\frac{1}{eR^{2}}\left[ \sqrt{\lambda \left( 1-\lambda 
\right) }\left( \left\langle 0_{1}^{+}\left| \hat{E}0 
\right| 0_{1}^{+}\right\rangle -\left\langle 0_{2}^{+} 
\left| \hat{E}0\right| 0_{2}^{+}\right\rangle
\right) \right.  \nonumber \\
&&\left. -\left( 2\lambda -1\right) \left\langle 0_{2}^{+} 
\left| \hat{E}0\right| 0_{1}^{+}\right\rangle \right]  
\label{rhobohr}
\end{eqnarray}
with $\hat{E}0=\sum_{i}e_{i}r_{i}^{2}.$ The crossed term 
in Eq. (\ref{rhobohr}) is strictly zero with either of the 
forces Sk3 or SG2 in the HF limit 
$\left( v_{i}^{2}=0,1\right) $ and can be neglected because 
of the large $\beta -$spacing in between the $0_{1}^{+}$ 
and $0_{2}^{+}$ minima in Fig. 1. One therefore gets for 
the $\rho ^{2}\left( E0\right) $ strength

\begin{equation}
\rho ^{2}\left( E0\right) =\lambda \left( 1-\lambda \right) 
\left[ \frac{Z}{R^{2}}\left( r_{1}^{2}-r_{2}^{2}\right) 
\right] ^{2}  \label{rho2}
\end{equation}
with $r_{1}$ and $r_{2}$ the charge r.m.s. radii of the 
$0_{1}^{+}$ and $0_{2}^{+}$ states. In principle, $R$ 
should be the r.m.s. radius of the ground state but since 
to our knowledge this has not been measured, we use 
$R=(r_{1}+r_{2})/2$ in Eq. (\ref{rho2}). An approximate 
expression for $\rho ^{2}\left( E0\right) $ in terms of 
$\beta _{2\left( 1\right) }$ and 
$\beta _{2\left( 2\right) }$ has been used in Ref. 
\cite{gelletly,heyde}.

\begin{equation}
\rho ^{2}\left( E0\right) =\lambda \left( 1-\lambda 
\right) \left[ \left( \beta _{2\left( 1\right) }^{2}-
\beta _{2 \left( 2\right) }^{2}\right) +
\frac{5\sqrt{5}}{21\sqrt{3}} \left( \beta _{2\left( 
1\right) }^{3}-\beta _{2\left( 2 \right) }^{3}\right) 
\right] ^{2}  \label{rho2bet}
\end{equation}

In Fig. 3 we plot $\rho ^{2}\left( E0\right) $ as a 
function of $\lambda $, obtained from both Eq. 
(\ref{rho2}) and Eq. (\ref{rho2bet}), using the r.m.s. 
and deformation values obtained with either Sk3 or SG2 
forces (see Table 2). One sees that using Eq. (\ref{rho2}) 
for Sk3 the experimental $E0$ value calls for a large 
mixing $\left( \lambda \left( \lambda -1\right)
\simeq 0.25\right) $ while for SG2 the mixing must be small 
$\left( \lambda \left( \lambda -1\right) \simeq 0.05\right) 
$. This is consistent with the fact that with SG2 the two 
minima differ more in energy and the barrier is higher than 
with Sk3 (see Fig. 1), hence a much smaller mixing 
probability is to be expected.

We would like to point out that also the $M1$ strength 
between the lowest $2^{+}$ states can be obtained in an 
analogous way from the gyromagnetic ratios $g_{R}$ of the 
$0_{1}^{+}$ and $0_{2}^{+}$ bands given in Table 2. The 
$B(M1)$ strength between the lowest $2^{+}$ states would 
be \cite{bohr}

\begin{equation}
B\left( M1,2_{II}^{+}\rightarrow 2_{I}^{+}\right) =
\frac{648}{4\pi }\lambda \left( 1-\lambda \right) 
\left[ g_{R,1}-g_{R,2}\right] ^{2}\mu _{N}^{2}
\label{bm12}
\end{equation}
For Sk3 with $\lambda =0.5$ and 
$g_{R,1}=0.487,\;g_{R,2}=0.459$ (see Table 2), we get 
$B\left( M1,2_{II}^{+}\rightarrow 2_{I}^{+}\right) 
=0.010\;\mu _{N}^{2}$ and for SG2 with $\lambda =0.05$ 
and $g_{R,1}=0.498,\;g_{R,2}=0.463 $ (see Table 2), we 
get $B\left( M1,2_{II}^{+}\rightarrow 2_{I}^{+}\right) 
=0.003\;\mu _{N}^{2}$.

\section{Gamow Teller strengths and sum rules}

The $\beta ^{+}$ strength distributions calculated in the 
selfconsistent HF+RPA scheme described in Section 2 are 
shown in Fig. 4 as a function of the excitation energy 
of the daughter nucleus. We present the results for both 
of the interactions considered, SG2 and Sk3, corresponding 
to $\beta ^{+}$ decay from each of the two minima, oblate 
and prolate. For comparison we also present the results 
that would correspond to $\beta ^{+}$ decay from the 
spherical shape. A strong shape dependence is observed in 
Fig. 4 that is roughly the same independent on whether the 
SG2 or the Sk3 interaction is used. The $Q_{EC}$ values 
are indicated by vertical lines. The $Q_{\beta }$ value 
for $\beta ^{+}$ decay has been calculated for each of 
the prolate, oblate, and spherical HF+BCS solutions as

\begin{equation}
Q_{\beta ^{+}}=M_{p}-M_{n}-m_{e}+\lambda _{p}- 
\lambda _{n}-E_{p}-E_{n}
\label{Qbeta}
\end{equation}
with $\lambda _{p}\left( \lambda _{n}\right) $ the proton 
(neutron) Fermi energy given in Table 2 and 
$E_{p}\left( E_{n}\right) $ the lowest proton (neutron) 
quasiparticle energy. The associated $Q-$value for electron
capture is

\begin{equation}
Q_{EC}=Q_{\beta ^{+}}+2m_{e}  \label{QEC}
\end{equation}
The values of $Q_{EC}$ obtained and shown in Fig. 4 are the 
following: $Q_{EC}=4.848$ MeV for the SG2 oblate solution, 
$Q_{EC}=4.628$ MeV for the SG2 prolate solution, 
$Q_{EC}=4.892$ MeV for the Sk3 oblate solution, and 
$Q_{EC}=4.308$ MeV for the Sk3 prolate solution. It is 
important to stress that the strength summed up to energies 
below the $Q_{EC}$ window is also strongly dependent on 
deformation (see also Fig. 8) and this should also be
experimentally observable.

Fig. 4 summarizes the main results of this section. One 
should keep in mind that a quenching of the $g_{A}$ factor 
$\left( g_{A,eff}=0.7g_{A}\right) $ is expected on the 
basis of the observed quenching in charge exchange reactions 
and spin $M1$ transitions in stable nuclei, where 
$g_{s,eff}$ is known to be approximately $0.7g_{s,free}$.

In what follows we disentangle the meaning of these results 
in the light of simpler approximations. In particular, we 
will study the sensitivity of the Gamow Teller strength 
distribution to RPA correlations and to pairing
correlations, as well as to the nuclear shape.

Since the distinct profiles of the strength distribution 
in the three nuclear shapes considered are not seriously 
modified from one interaction to another we restrict the 
discussion in what follows to results obtained with the 
SG2 interaction.

In the laboratory frame the transition probability for 
$\beta ^{+}$ decay from the $0^{+}$ to a $1_{\omega }^{+}$ 
state is given by

\begin{equation}
B_{GT}^{+}\left( 0^{+}\rightarrow 1_{\omega }^{+}\right) = 
\frac{g_{A}^{2}}{4\pi }\left| \left\langle 1_{\omega }^{+} 
\left\| \beta ^{+}\right\| 0^{+}\right\rangle \right| ^{2}  
\label{BGT}
\end{equation}
and the reduced matrix element $\left\langle 1_{\omega }^{+} 
\left\| \beta ^{+}\right\| 0^{+}\right\rangle $ in the 
laboratory frame is related to the intrinsic matrix elements 
$\left\langle \omega _{K}\left| \beta _{K}^{+}\right| 0 
\right\rangle $ discussed in the previous section by

\begin{eqnarray}
\left| \left\langle 1_{\omega }^{+}\left\| \beta ^{+} 
\right\| 0^{+}\right\rangle \right| ^{2} &=&2\left| 
\left\langle \omega _{1}\left| \beta _{1}^{+}\right| 0 
\right\rangle \right| ^{2},\;\omega =\omega _{1} 
\nonumber \\
&=&\left| \left\langle \omega _{0}\left| \beta _{0}^{+} 
\right| 0\right\rangle \right| ^{2},\;\omega =\omega _{0}  
\label{intrin}
\end{eqnarray}
where we have used the Bohr-Mottelson factorization 
approximation \cite{bohr} and have neglected possible 
higher order corrections due to angular momentum projection 
\cite{moya}. Therefore, the calculated strength functions 
can be written as

\begin{equation}
B_{GT}^{+}\left( \omega \right) =\frac{g_{A}^{2}}{4\pi } 
\left\{ \sum_{\omega _{0}}\left| \left\langle \omega _{0} 
\left| \beta _{0}^{+}\right| 0\right\rangle \right| ^{2} 
\delta \left( \omega -\omega _{0}\right) +2 \sum_{\omega 
_{1}}\left| \left\langle \omega _{1}\left| \beta _{1}^{+}
\right| 0\right\rangle \right| ^{2}\delta \left( \omega 
-\omega _{1}\right) \right\}  \label{bstrength}
\end{equation}
consisting of spikes at the various 
$\omega _{K}\;\left( K=0,1\right) $ solutions.

To facilitate the comparison among the various 
approximations discussed here we have folded the calculated 
GT strengths with $\Gamma =1$ MeV width Gaussians converting 
the discrete spectrum into a continuous profile. This can 
also be interpreted as an approximate way to incorporate 
the further fragmentation and smoothing effects due to 
coupling to other excitation modes not taken into account 
in RPA. The effect of this folding procedure is illustrated 
in Fig. 5 where we compare the strength in the discrete 
spectrum to the convoluted strength distribution for the 
particular case of the prolate solution with the SG2 
interaction. In the lower panels the $K=0$ and $K=1$ 
strengths are represented by vertical dotted and solid 
lines, respectively. The total strength in the upper parts 
of Fig. 5 is represented by a solid line. The right panels 
contain the results of the RPA calculations while the left 
panels contain the results of calculations for $\beta ^{+}$ 
decay to uncorrelated neutron proton two quasiparticle 
$\left( 2qp\right) $ excitations. The large difference 
observed between the results of RPA and uncorrelated 2qp 
neutron-proton excitations in the discrete spectra in 
Fig. 5 is reliably reflected in the folded distribution. 
Hence, from here on we restrict our presentation to folded 
strength distributions.

\subsection{Comparison of RPA, TDA, and bare 2qp strengths}

The importance of the role played by the residual 
interaction is illustrated in Fig. 6 where we compare 
results for GT strength distributions obtained in RPA 
(solid line), in TDA (dashed line), and in the uncorrelated
two-quasiparticle limit (dotted line). Clearly the repulsive 
spin-isospin residual interaction $\left( V_{GT}\right) $ 
moves the strength to higher energy. This effect is already 
present when one goes from the bare 2qp limit to the Tamm 
Dancoff approximation. The deformation dependence of the
strength agrees with that found by Hamamoto and 
collaborators \cite{hamamoto} in the context of TDA. The 
main effect of the RPA correlations, that can be seen 
comparing TDA to RPA, is to provide a reduction of the TDA 
strength maintaining the position of the peaks practically 
unchanged. A strong dependence of the GT strength on the 
shape of the $\beta ^{+}$ parent is observed in the three 
approximations, being more dramatic in the uncorrelated 
situation. This reflects the fact that the dependence on 
the different internal shell structure involved in the 
different shapes is stronger in the absence of a residual 
interaction. The latter tends to redistribute the strength 
and tends to produce a smoother strength distribution. 
This is further illustrated in Fig. 7 where we show the RPA
results for oblate, spherical, and prolate shapes 
corresponding to different coupling strengths of the 
$V_{GT}$ interaction. The results corresponding to 
$\chi _{GT}=0.48$ in Table 1 are compared to those 
obtained when the strength $\chi _{GT}$ of the interaction 
is reduced or increased by a factor of 2.

The summed strengths up to different excitation energies 
are represented in Fig. 8 for the two minima (oblate and 
prolate) corresponding to the SG2 interaction. The results 
with Sk3 are similar. Also plotted in the figure are the 
linear energy-weighted sum strengths. Clearly the 
uncorrelated sum strength

\begin{equation}
\sum_{x}\left( B_{x}^{GT\;+}\right) _{uncorr}= 
\frac{g_{A}^{2}}{4\pi }\sum_{\alpha \beta }\left| 
\left\langle \alpha \beta \left| \bbox{\sigma}t^{+} 
\right| \phi _{0}\right\rangle \right| ^{2}  \label{sumb}
\end{equation}
is conserved by the TDA approximation while the RPA 
approximation conserves the uncorrelated LEWSR

\begin{equation}
\sum_{x}\left( E_{x}B_{x}^{GT\;+}\right) _{uncorr}= 
\frac{g_{A}^{2}}{4\pi }\sum_{\alpha \beta } \left( 
E_{\alpha }+E_{\beta }\right) \left| \left\langle \alpha 
\beta \left| \bbox{\sigma}t^{+}\right| \phi _{0}
\right\rangle \right| ^{2}  \label{sumeb}
\end{equation}
where $\alpha \beta $ represents any two quasiparticle 
neutron-proton excitation. Although this is to be expected 
from general theory, it is interesting to see that already 
at low energies ($E^{*}\lesssim 10$ MeV) in RPA the linear 
energy-weighted sum tends to converge to the bare 2qp 
result, while the non energy-weighted sum is clearly lower 
than the bare 2qp result in order to compensate for the 
push to higher energies of the strength produced by the 
$V_{GT}$ interaction. The opposite happens with TDA. The
lack of RPA correlations makes the TDA linear 
energy-weighted sum to start diverging already at 10 MeV. 
Also shown in Fig. 8 and indicated by arrows, are the 
calculated $Q_{EC}$ values for the oblate and prolate 
shapes where one sees that the predicted summed 
$\beta ^{+}$ strength below $Q_{EC}$ is larger in the 
prolate case. The total linear energy-weighted and non
energy-weighted sums up to an excitation energy of 30 MeV 
are given in Table 3. The results correspond to 
$\beta ^{+}$ decay from the two minima (prolate and 
oblate) with the SG2 interaction. Also given for 
comparison are the results for $\beta ^{+}$ decay from 
the prolate minimum with Sk3 interaction. Here one can 
also see (see Fig. 4) that the results depend more on 
the shape of the parent nucleus than on the effective 
interaction. For completeness in Table 3 we give also 
the $\beta ^{-}$ summed strength and in the column 
labelled Ikeda, we give the difference between 
$\beta ^{-}$ and $\beta ^{+}$ summed strengths to show 
that the Ikeda sum rule is satisfied in all cases with 
an accuracy of $0.3\%$. The summed strengths for $K=0$ 
and $K=1$ are given separately to check that Ikeda's sum 
rule is satisfied independently for each $K$ value.

\subsection{Analysis of pairing and deformation effects}

As already stated our mean field calculations only include 
neutron-neutron and proton-proton pairing correlations. 
Taking into account neutron-proton pairing would increase 
the diffuseness of the Fermi surface (see Goodman 
\cite{nppairing}), which is governed by the gap parameters. 
It is therefore interesting to study the sensitivity of 
the GT strength to this diffuseness. To this end we 
compare in Fig. 9 the RPA results obtained with the SG2 
force for different values of the gap parameters 
$\Delta _{n}=\Delta _{p}=1.5$ MeV, 
$\Delta _{n}=\Delta _{p}=1.0$ MeV, and 
$\Delta _{n}=\Delta _{p}=0.5$ MeV. In the spherical case 
the role of pairing is different in the low 
($E^{*}\simeq 1$ MeV) and high ($E^{*}\simeq 6$ MeV) peaks. 
The first peak decreases with increasing pairing while the 
second peak increases. In the deformed cases (oblate and 
prolate) the role of pairing is somewhat less pronounced, 
but as a general rule, we see that with increasing pairing 
the strength at high energies (beyond $E^{*}\simeq 4$ MeV) 
increases.

To understand better the role of pairing and deformation we 
show in Fig. 10 the results of bare two quasiparticle 
excitations with and without pairing for the various shapes. 
From left to right we can see the effect of going from 
prolate to spherical and to oblate. From down to upper 
panels we can see the effect of going from no pairing 
$\left( \Delta =0\right) $ to pairing ($\Delta =1.5$ MeV). 
In these figures the solid lines (both in the discrete and 
continuous distributions) correspond to $K=1$ excitations 
and the dotted lines to $K=0$ excitations. When there is 
no pairing and no deformation (Fig. 10e) there is only one 
neutron-proton particle-hole channel open 
$\left( \pi p_{3/2}\rightarrow \nu p_{1/2}\right) $. When 
we include pairing maintaining the spherical shape 
(Fig. 10b), the strength of this channel decreases and new 
channels are open, in particular the 
$\pi f_{7/2}\rightarrow \nu f_{5/2}$ that is forbidden in 
the case $\Delta =0,\beta =0$ (see also Fig. 11). 
Consequently a two peak structure appears and it remains 
in the spherical RPA results (see for instance Fig. 4), 
even though it is modulated by the effect of the residual 
interaction and RPA correlations. Deformation alone also 
causes the opening of new channels and the reduction of 
the $\pi p_{3/2}\rightarrow \nu p_{1/2}$ peak (see the
transition from panel (e) to panels (d) and (f) in Fig. 10), 
but in addition it causes fragmentation of the strength 
between single particle spherical orbitals. In particular, 
the $K=0$ and $K=1$ modes are degenerate in the spherical 
case independently of whether there is pairing or not. This
degeneracy is clearly broken in the deformed case: only 
$K_{\pi }^{\pm }\rightarrow K_{\nu }^{\pm }$ transitions 
with $K_{\pi }=K_{\nu }=1/2$ contribute to both $K=0$ and 
$K=1$ modes, while $K_{\pi }^{\pm }\rightarrow 
K_{\nu }^{\pm }$ transitions with $K_{\nu }=K_{\pi }\pm 1\ 
\left( K_{\nu },K_{\pi }>1/2\right) $ contribute only to 
$K=1$ modes, and those with $K_{\nu }=K_{\pi }>1/2$ 
contribute only to $K=0$ modes. Clearly which are the new 
channels that are open with deformation, and what are their 
strengths, depends strongly on whether the nucleus is 
oblate or prolate and on the magnitude of the deformation. 
As seen in Fig. 10d (see also Fig. 11) in the prolate 
$\Delta =0$ case the strongest new channel is the $K=1\ 
\left( 7/2_{\pi }^{-}\rightarrow 5/2_{\nu }^{-}\right) $, 
that has a strong spherical component $\pi f_{7/2}
\rightarrow \nu f_{5/2}$, while the $\pi p_{3/2}
\rightarrow \nu p_{1/2}$ transition is negligibly small, 
and other positive parity states have also more strength. 
On the contrary in the oblate case the strength is still 
due to the $\pi p_{3/2}\rightarrow \nu p_{1/2}$ transition 
that appears strongly fragmented into several $3/2_{\pi
}^{-}\rightarrow 1/2_{\nu }^{-}$ and $1/2_{\pi }^{-}
\rightarrow 1/2_{\nu }^{-}$ transitions. The effect of 
pairing in the deformed case is similar to that in the 
spherical case. It opens more channels and reduces the 
strength of the existing ones (compare panels (a) to (d) 
and (c) to (f) in Fig. 10), resulting in more complex 
and smoother strength distributions. It is also interesting 
to observe that in spite of the redistributions of strength
caused by the residual interaction $V_{GT}$ and RPA 
correlations, the distinctive prolate, spherical, and oblate 
profiles of the $B_{GT}$ strengths in Fig. 4 are still 
reminiscent of those in Fig. 10. Hence, the above 
discussion helps to trace back the origin of the main 
features observed in Fig. 4.

In order to understand better the mechanisms leading to 
such patterns in the profiles of the GT strength 
distribution it is illustrative to compare the HF single 
particle neutron and proton spectra around the Fermi levels 
in the spherical and deformed cases. This can be seen in 
Fig. 11, where the Fermi energies are indicated by the step 
dashed lines. The labels correspond to the $\ell _{j}$ 
values in the spherical case and to $K^{\pi }$ in the
deformed cases. In the absence of pairing, the allowed 
GT$^{+}$ transitions connect only protons below 
$\lambda _{p}$ with neutrons above $\lambda _{n}$ with 
$\Delta K=0,\pm 1$ and same parity. The most important GT 
transitions are indicated by the solid arrows. They are 
$(\pi p_{3/2}\rightarrow \nu p_{1/2})$ in the spherical 
case, $(\pi 7/2^{-}\rightarrow \nu 5/2^{-})$ in the 
prolate case, and $(\pi 1/2^{-}\rightarrow \nu 1/2^{-})$ 
in the oblate one. The dashed arrow indicates the most 
important GT transition in the spherical case 
$\left( \pi f_{7/2}\rightarrow \nu f_{5/2}\right) $, which 
is allowed once pairing correlations are included. From 
this figure one can also understand why the GT strength in 
the oblate case is much smaller than in the prolate one. 
As seen in Fig. 11 the reason for this behaviour is that
in the oblate case the proton levels below the Fermi 
energy are negative parity states while those above the 
neutron Fermi level are mostly positive parity states, 
while in the prolate case one finds both, positive and
negative parity states, below $\lambda _{p}$ and above 
$\lambda _{n}$. This figure can also be used to predict 
behaviours of GT strengths when we add or subtract 
nucleons to $^{74}$Kr. For instance, when we remove two 
neutrons, the state $\nu 3/2^{-}$ in the oblate case will 
appear above the Fermi level and therefore it can be fed 
by $\beta ^{+}$ decay of the negative parity proton 
states below the proton Fermi level. Thus, one expects an 
important increase of the GT strength when going from the 
oblate $^{74}$Kr to the oblate $^{72}$Kr. On the contrary, 
based on the same picture, we do not expect much changes 
in the prolate case.

\subsection{Magnetic properties}

Gyromagnetic factors and estimates of the total $M1$ 
transition strength between the lowest $2^{+}$ states were 
given at the end of Section 3. In this Section we focus 
on $0^{+}\rightarrow 1^{+}$ spin excitations, which
are stronger.

The spin $M1$ transitions are the $\Delta T_{z}=0$ isospin 
counterparts of the $\Delta T_{z}=\pm 1$ Gamow Teller 
transitions. The study carried out in \cite{sar96} for 
several isotope chains showed that the $M1$ strength
depends on deformation, not only for orbital but also for 
spin excitations. Thus, on general grounds, one may argue 
that the deformation dependence of the GT strength 
discussed earlier should be comparable to that of the spin 
$M1$ strength considered in \cite{sar96}.

To see to what extent this argument holds we show in Fig. 
12 the profiles of the spin $M1$ strength distributions for 
the three shapes oblate, spherical, and prolate in 
$^{74}$Kr. The results correspond to selfconsistent HF+RPA
calculations with the SG2 interaction. These calculations 
are described in \cite{sar96,sar97} and closely follow the 
description of calculations in Section 3 except for the 
$\Delta T_{z}=0$ character of the $M1$ operator that, in 
particular, demands orthogonality of the RPA\ states to the
spurious rotational state \cite{noj}. For completeness we 
also show the results obtained in the absence of residual 
spin-spin interaction ($K_{S}=0$ in Eq. (\ref{hss})), i.e., 
the results for bare two quasiparticle spin $M1$ 
excitations.

One can see in Fig. 12 that as in the case of GT 
excitations (compare with Fig. 6) the residual interaction 
produces a redistribution of the strength pushing it to 
higher energies. There is a striking similarity between the
shape dependence in Fig. 12 and that in Fig. 6. This is so 
in both 2qp and RPA strength distributions.

Although it is unlikely that $M1$ transitions may be 
observed in such highly unstable nuclei, the comparison of 
Figs. 6 and 12 tells us that the main features of GT and 
spin $M1$ strength distributions are very similar. This
suggests that we may learn about properties that are 
observable in highly unstable nuclei (like $\beta $ decay) 
from properties that are observable in stable nuclei 
(like $M1$'s), and vice versa.

\section{Total half-life}

Another quantity of interest in nucleosynthesis is the 
$\beta ^{+}/EC$ half-lives of exotic proton rich nuclei. 
The total half-life depends on the strength distribution 
in a different way from both the energy-weighted and
non-weighted sums considered above. It is therefore 
illustrative to see the predictions for the half-life as 
well as the sensitivity of this quantity to the various 
effects considered in this work.

The partial half-life $t_{1/2}$ for a given transition 
connecting the ground state of a nucleus with an excited 
state in the daughter nucleus is given by

\begin{equation}
f\left( Z,\omega \right) t_{1/2}=\frac{D}{\kappa ^{2}\left| 
\left\langle 1_{\omega }^{+}\left\| \beta ^{+}\right\| 
0^{+}\right\rangle \right| ^{2}}
\label{ftparcial}
\end{equation}
where we have neglected the Fermi strength. 
$f\left( Z,\omega \right) $ is the Fermi integral 
\cite{astro,gove}. We use $D=6200$ s and include
effective factors

\begin{equation}
\kappa ^{2}=\left[ \left( g_{A}/g_{V}\right) _{eff}
\right] ^{2}=\left[ 0.7\left( g_{A}/g_{V}\right) _{free}
\right] ^{2}=0.74  \label{geff}
\end{equation}

The total half-life $T_{1/2}$ for allowed $\beta $ decay 
from the ground state of the parent nucleus is given by 
summing over all the final states involved in the process

\begin{equation}
T_{1/2}^{-1}=\frac{0.74}{6200}\sum_{\omega }f\left( Z,
\omega \right) \left| \left\langle 1_{\omega }^{+}\left\| 
\beta ^{+}\right\| 0^{+}\right\rangle \right| ^{2}  
\label{totalt}
\end{equation}
The Fermi integrals $f\left( Z,\omega \right) $ are taken 
from Ref. \cite {gove}.

One should notice that the sum over final states in Eq. 
(\ref{totalt}) is extended over all the states in the daughter 
nucleus with excitation energies below the $Q-$value. 
Another difference with respect to the energy-weighted and 
non-weighted sums considered earlier is that in the sum
of Eq. (\ref{totalt}) the weighting factor is larger at low 
excitation energies, which correspond to large kinetic 
energies of the emitted positron.

We can see in Table 4 the results obtained from bare $2qp$, 
TDA, and RPA calculations for the total $\beta ^{+}/EC$ 
half-life of $^{74}$Kr. Results are shown for the two 
Skyrme forces Sk3 and SG2, as well as for the two shapes 
oblate and prolate. The experimental value is 
$\left( T_{1/2}\right) _{EXP}=11.5$ min. It is clear that 
the half-lives calculated with both Sk3 and SG2 increase 
going from $2qp$ to RPA and are larger in the prolate case.
On the other hand, the oblate SG2 produces larger $T_{1/2}$ 
values than the oblate Sk3, while the prolate SG2 gives 
smaller $T_{1/2}$ values than the corresponding prolate Sk3.

Although not shown in the table we have also checked the 
sensitivity of this quantity to other effects considered 
in this work such as the strength of the residual 
interaction $\chi _{GT}$ and the pairing gaps. For example,
from an RPA calculation with the force SG2 we find that 
changing the strength $\chi _{GT}$ in Table 1 to half or 
twice its value, $T_{1/2}$ changes from 3.5 to 16.7 min in 
the oblate case and from 6.9 to 23.8 min in the prolate 
case, i.e., roughly speaking, within RPA, $T_{1/2}$ 
increases proportionally to $\chi _{GT}$. The results are 
also sensitive to the pairing gaps. Again, an RPA 
calculation with SG2 and pairing gaps 
$\Delta _{n}=\Delta _{p}=1$ MeV, gives $T_{1/2}=1.8$ min 
and $T_{1/2}=4.9$ min in the oblate and prolate cases, 
respectively, i.e., the half-life decreases with 
decreasing pairing.

One should also keep in mind that values for the effective 
$g_{A}$ factors ranging from 0.7 to 0.8 of the bare values 
are considered in the literature. Since we have taken here 
the value 0.7, the half-lives in Table 4 would be reduced 
a 20\% by using the value 0.8.

\section{Summary and final remarks}

We have investigated shape isomerism and $\beta $ decay in 
$^{74}$Kr on the basis of the selfconsistent HF+RPA 
framework with Skyrme forces. This is a well founded method 
that has been successfully used to describe quite diverse 
properties of stable spherical and deformed nuclei over the 
periodic table. The merits of this approximation are well 
known \cite{sar96}-\cite{bertsch}. Once the effective 
interaction parameters are determined by fits to global 
properties in spherical nuclei over the periodic table and 
the gap parameters of the pairing force are specified, 
there are no free parameters left. Both the residual 
interaction and the mean field are consistently obtained 
from the same two-body interaction. This is particularly 
desirable for nuclei far from stability, where extrapolation 
of methods based on phenomenological mean fields are more 
doubtful. With the method used here one predicts ground 
state and low-lying properties, as well as $\beta $ decay 
and spin $M1$ excitations without introducing any free 
parameter. We obtain a fair description of the 
experimentally known properties of $^{74}$Kr and, in 
particular, of the interesting shape isomerism. In this 
regard one question is open as to whether the ground 
state is prolate and the first excited $0^{+}$ state is 
oblate or vice versa, since different Skyrme interactions 
give different answers. All the Skyrme interactions that we
tried (Sk3, SG2, Ska, and Skm$^{\star }$) give a similar 
stable prolate shape $\left( \beta _{2}\simeq 0.4\right) $, 
but their predictions on the stable oblate shape are 
somewhat different ranging from $\beta \simeq -0.15$ up to 
$\beta \simeq -0.30$. These deformations are compatible with 
the experimental $B(E2)$ \cite{heese} and 
$\rho ^{2}\left( E0\right) $ \cite{gelletly} values. 
Theoretical predictions for gyromagnetic ratios and $M1$
transitions are also given.

The RPA Gamow Teller $\beta ^{+}$ strength distributions 
depends strongly on the shape (prolate, spherical or oblate) 
of the $\beta ^{+}$ parent $^{74}$Kr, and remarkably, this 
dependence is quite noticeable in the experimentally 
accessible energy window. A close similarity is found 
between this deformation dependence and the deformation 
dependence of the spin $M1$ strength distribution. It is 
important to emphasize that these results do not depend 
much on which effective Skyrme interaction (Sk3 or SG2) 
is used.

We have also studied the dependence of the results on 
various aspects of the theory, in particular on the 
residual interaction, RPA, and pairing correlations. 
Compared to the uncorrelated two quasiparticle response, 
RPA shifts the GT strength to higher energies and reduces 
the total strength. These effects are of course more 
pronounced if one increases the coupling strength of the 
residual interaction. While the shifting effect is already
contained in the TDA description, the quenching effect is 
not. This is consistent with the fact that the non 
energy-weighted summed strength is conserved in TDA, while 
the linear energy weighted summed strength is conserved in 
RPA. BCS correlations reduce the strength of allowed
particle-hole excitations and create new ones; the main 
effect of increasing the Fermi diffuseness is to smooth 
out the profile of the GT $\beta ^{+}$ strength 
distribution increasing the strength at high energies.

Predictions for the half-life are given together with a 
discussion of the sensitivity of this quantity to the 
various effects considered in this work. The experimental 
value, $\left( T_{1/2}\right) _{EXP}=11.5$ min, is in
between the two RPA results for the prolate solutions 
of the SG2 and Sk3 forces.

It will be interesting to test our results with 
experimentally accessible information on GT strengths. 
Since results on the proton rich odd-A Kr isotopes will 
be soon available \cite{miehe}, it will also be interesting 
to extend our calculations to include odd-A nuclei. 
Comparison to data would be desirable before considering 
further theoretical refinements. The latter would deal 
with issues such as extended RPA, or the inclusion of the
continuum and neutron-proton pairing at the mean field 
level. It is however important to keep in mind that in 
the present method the Ikeda sum rule is conserved, 
while violations of this sum rule are found in some of 
these extensions.

\acknowledgments 

We are thankful to J. Dukelsky, A. Faessler, M.J. 
Garc\'\i a Borge, W. Gelletly, and Ch. Mieh\'{e} for 
stimulating comments and discussions. This work was 
supported by DGICYT (Spain) under contract number 
PB95/0123 and by the EC program 'Human Capital and 
Mobility' under Contract No. CHRX-CT 93-0323. One of 
us (A.E.) thanks Ministerio de Educaci\'{o}n y Cultura
(Spain) for support.

\newpage

\begin{figure}

{\bf Figure 1. }Total energy as a function of the mass 
quadrupole moment obtained from a constraint HF+BCS 
calculation with the Skyrme interaction SG2 (solid) 
and Sk3 (dashed).

\end{figure}

\begin{figure}

{\bf Figure 2. }Experimental spectrum for $^{74}$Kr 
\cite{tabor} compared to the theoretical rotational 
spectra obtained for the oblate and prolate solutions 
of the SG2 and Sk3 interactions.

\end{figure}

\begin{figure}

{\bf Figure 3. }$E0$ strength from Eq. (\ref{rho2}) for 
SG2 (solid line) and for Sk3 (dotted line). Also shown 
is the $E0$ strength from Eq. (\ref{rho2bet}) for SG2 
(dashed line) and for Sk3 (dash-dotted line). The strength 
is plotted as a function of the mixing parameter $\lambda $ 
(see text). The experimental value from \cite{gelletly} 
is also indicated.

\end{figure}

\begin{figure}

{\bf Figure 4.} Gamow Teller strength distribution in 
$^{74}$Kr as a function of the excitation energy of the 
daughter nucleus. The results correspond to the forces 
SG2 (solid lines) and Sk3 (dashed lines) in RPA with 
the coupling strengths given in Table 1. The results 
are for the prolate (upper part), spherical (middle 
part), and oblate (lower part) shapes. The vertical 
solid (dashed) lines indicate the SG2 (Sk3) $Q_{EC}$
values.

\end{figure}

\begin{figure}

{\bf Figure 5. }Energy distribution of the Gamow Teller 
strength for the prolate solution of the force SG2 in 
$^{74}$Kr. The lower panels show the spectrum of $K=0$ 
(dashed vertical lines) and $K=1$ (solid vertical lines)
excitations. The upper panels show the corresponding 
folded strength distributions (see text) for $K=0$ (short 
dashed line), $K=1$ (long dashed line), and total (solid 
line) GT strength. The left (right) panels are from bare 
two quasiparticle (RPA) calculations.

\end{figure}

\begin{figure}

{\bf Figure 6. } Comparison of RPA (solid line), TDA 
(dashed line), and bare two quasiparticle (dotted line) 
GT strength distributions in $^{74}$Kr. The results 
correspond to the force SG2 for the prolate, spherical, 
and oblate solutions. Also shown with vertical lines 
are the $Q_{EC}$ values.

\end{figure}

\begin{figure}

{\bf Figure 7.} RPA\ Gamow Teller strength distributions 
in $^{74}$Kr for different coupling strengths of the 
residual interaction $\chi _{GT}$. The solid lines 
correspond to the value in Table 1. The short (long) dashed
lines correspond to a half (double) value of the coupling 
strength.

\end{figure}

\begin{figure}

{\bf Figure 8.} Accumulated GT strength 
$\left( \Sigma B\;\left[ g_{A}^{2}/4\pi \right] \right) $ 
and energy-weighted strength $\left( \Sigma EB\;\left[ 
MeV\;g_{A}^{2}/4\pi \right] \right) $ in $^{74}$Kr as a 
function of the excitation energy of the daughter nucleus. 
The results correspond to the prolate (upper part) and 
oblate (lower part) solutions of SG2. Dotted lines are 
for bare 2qp calculations, dashed lines for TDA 
calculations, and solid lines for RPA calculations. The 
vertical lines indicate the $Q_{EC}$ values.

\end{figure}

\begin{figure}

{\bf Figure 9.} Pairing effect in the RPA Gamow Teller 
strength distribution in $^{74}$Kr. The solid lines 
correspond to calculations with pairing gaps 
$\Delta _{n}=\Delta _{p}=$ 1.5 MeV, long (short) dashed 
lines are for $\Delta _{n}=\Delta _{p}=$ 1.0 MeV 
($\Delta _{n}=\Delta _{p}=$ 0.5 MeV). As in previous 
figures we can see the results for the prolate (upper 
part), spherical (middle part), and oblate (lower part) 
solutions.

\end{figure}

\begin{figure}

{\bf Figure 10. }Comparison of pairing and deformation 
effects in the Gamow Teller strength distributions of 
$^{74}$Kr. The results correspond to bare two 
quasiparticle calculations with the force SG2. From left 
to right we can see the prolate, spherical, and oblate 
cases. The upper panels correspond to results with 
pairing $\left( \Delta _{n}=\Delta _{p}=1.5MeV\right) $ 
and the lower ones are without pairing 
$\left( \Delta =0\right) $. Solid lines (spectra and folded 
distributions) correspond to $K=1$ excitations, while
dashed lines (spectra and folded distributions) correspond 
to $K=0$ excitations. The main configurations leading to 
the GT excitations are also shown.

\end{figure}

\begin{figure}

{\bf Figure 11. }Spherical, prolate, and oblate single 
particle energies for protons and neutrons obtained from 
HF calculations with the force SG2. The Fermi energies 
are indicated by the step dashed lines. The most 
important GT transitions in the spherical and deformed 
cases are indicated by the solid arrows while the dashed 
arrow indicates the most important GT transition in the 
spherical case allowed once pairing correlations are included.

\end{figure}

\begin{figure}

{\bf Figure 12.} Energy distribution of the spin $M1$ 
strength with $K^{\pi }=1^{+}$ in $^{74}$Kr. The results 
correspond to the force SG2 and pairing gap parameters 
$\Delta _{n}=\Delta _{p}=$ 1.5 MeV. Dashed lines are for 
2qp calculations and solid lines for RPA calculations. 
The three panels are for the prolate (upper part), 
spherical (middle part), and oblate (lower part) solutions.

\end{figure}

\newpage

\begin{table}

{\bf Table 1. }Parameters of the Skyrme forces SG2 and Sk3: 
$t_{0}$ $\left[ \text{MeV fm}^{3}\right] $, $t_{1}$ $\left[ 
\text{MeV fm}^{5}\right] $, $t_{2}$ $\left[ 
\text{MeV fm}^{5}\right] $, $t_{3}$ $\left[ 
\text{MeV fm}^{6}\right] $, $W$ $\left[ \text{MeV fm}^{5} 
\right] $, $x_{0}$, $x_{1}$, $x_{2}$, $x_{3}$, and 
$\alpha $. Also shown are the strengths of the separable 
isospin $\chi _{F}\;\left[ \text{MeV}\right] $ and 
spin-isospin $\chi _{GT}$ $\left[ \text{MeV}\right] $ 
residual interactions obtained from
Eqs.(\ref{X_F}) and (\ref{X_GT}) , respectively.

\begin{tabular}{ccccccccccccc}
& $t_{0}$ & $t_{1}$ & $t_{2}$ & $t_{3}$ & $W$ & $x_{0}$ 
& $x_{1}$ & $x_{2}$ & $x_{3}$ & $1/\alpha $ & $\chi _{F}$ 
& $\chi _{GT}$ \\ 
SG2 & $-2645.0$ & $340.0$ & $-41.9$ & $15595.0$ & $105.0$ 
& $0.09$ & $-0.0588 $ & $1.425$ & $0.06044$ & $6.0$ 
& $0.69$ & $0.48$ \\ 
Sk3 & $-1128.75$ & $395.0$ & $-95.0$ & $14000.0$ & $120.0$ 
& $0.45$ & $0.0$ & $0.0$ & $1.0$ & $1.0$ & $0.88$ & $0.46$
\end{tabular}

\end{table}

\vskip .5cm

\begin{table}

{\bf Table 2. } Fermi energies [MeV] for neutrons 
$\left( \lambda _{n}\right) $ and protons 
$(\lambda _{p}),$values of 
$\left\langle J^{2}\right\rangle $, cranking moments 
of inertia ${\cal I}_{cr}$, gyromagnetic ratios 
$g_{R}$, charge radii $r_{c}$, deformation parameters 
$\beta _{2}$, total energies $E_{HF}$, and $0^{+}$ 
energies from Eq. (\ref{E0+}) for the two minima 
(oblate and prolate) of the two interactions, Sk3 and SG2.

\begin{tabular}{ccccccccccc}
&  & $\lambda _{n}$ & $\lambda _{p}$ 
& $\left\langle J^{2}\right\rangle $ 
& ${\cal I}_{cr}$ & $g_{R}$ & $r_{c}$ & $\beta _{2}$ 
& $E_{HF}$ & $E_{0+}$ \\ 
Sk3 & oblate & -12.300 & -3.797 & 21.5 & 3.6 & 0.487 
& 4.229 & -0.259 & -625.40 & -628.4 \\ 
& prolate & -12.589 & -4.290 & 60.3 & 8.9 & 0.459 
& 4.269 & 0.386 & -625.89 & -629.3 \\ 
SG2 & oblate & -12.739 & -3.901 & 13.4 & 2.4 & 0.498 
& 4.147 & -0.147 & -643.92 & -646.7 \\ 
& prolate & -12.752 & -4.139 & 60.2 & 9.1 & 0.463 & 4.230 
& 0.389 & -642.87 & -646.2
\end{tabular}
{\bf \ }

\end{table}

\vskip .5cm

\begin{table}

{\bf Table 3. }Results obtained from bare 2qp, TDA, and 
RPA calculations corresponding to the oblate and prolate 
solutions of SG2 and to the prolate solution of Sk3 in 
$^{74}$Kr. The third (fourth) column contains the GT
strength in units of [$g_{A}^{2}/4\pi $] of the 
$\beta ^{-}\left( \beta ^{+}\right) $ decay summed up 
to $E_{cut}=30$ MeV. The fifth column contains the Ikeda 
sum rule obtained for this energy cut and the last column 
contains the energy-weighted summed strength in units of 
[$MeV\;g_{A}^{2}/4\pi $] corresponding to the $\beta ^{+}$ 
decay. In all cases the results are for $K=0$ and $K=1$ 
(within parentheses).

\begin{tabular}[t]{cccccc}
&  & $\sum B_{GT}^{-}$ & $\sum B_{GT}^{+}$ & Ikeda 
& $\sum EB_{GT}^{+}$ \\ 
SG2 oblate &  &  &  &  &  \\ 
& 2qp & 4.517 (8.305) & 2.523 (4.315) & 5.984 
& 12.56 (23.92) \\ 
& TDA & 4.517 (8.305) & 2.523 (4.315) & 5.984 
& 16.95 (31.22) \\ 
& RPA & 3.880 (7.228) & 1.886 (3.239) & 5.984 
& 12.65 (23.76) \\ 
SG2 prolate &  &  &  &  &  \\ 
& 2qp & 5.294 (11.952) & 3.300 (7.964) & 5.983 
& 17.32 (35.59) \\ 
& TDA & 5.294 (11.952) & 3.300 (7.964) & 5.983 
& 23.56 (52.32) \\ 
& RPA & 4.431 (9.758) & 2.437 (5.769) & 5.983 
& 17.31 (36.87) \\ 
Sk3 prolate &  &  &  &  &  \\ 
& 2qp & 5.389 (12.079) & 3.395 (8.088) & 5.986 
& 18.86 (37.84) \\ 
& TDA & 5.389 (12.079) & 3.395 (8.088) & 5.986 
& 25.28 (54.93) \\ 
& RPA & 4.533 (9.888) & 2.538 (5.898) & 5.986 
& 18.79 (39.03)
\end{tabular}

\end{table}

\vskip .5cm

\begin{table}

{\bf Table 4. } Total half-lives $T_{1/2}$ [minutes] 
of $^{74}$Kr.

\begin{tabular}{cccccccc}
& oblate Sk3 &  & oblate SG2 &  & prolate SG2 &  
& prolate Sk3 \\ 
&  &  &  &  &  &  &  \\ 
$2qp$ & 0.7 &  & 1.3 &  & 2.8 &  & 3.7 \\ 
TDA & 3.3 &  & 5.2 &  & 7.6 &  & 13.9 \\ 
RPA & 4.3 &  & 6.9 &  & 9.8 &  & 16.7
\end{tabular}
{\bf \ }

\end{table}

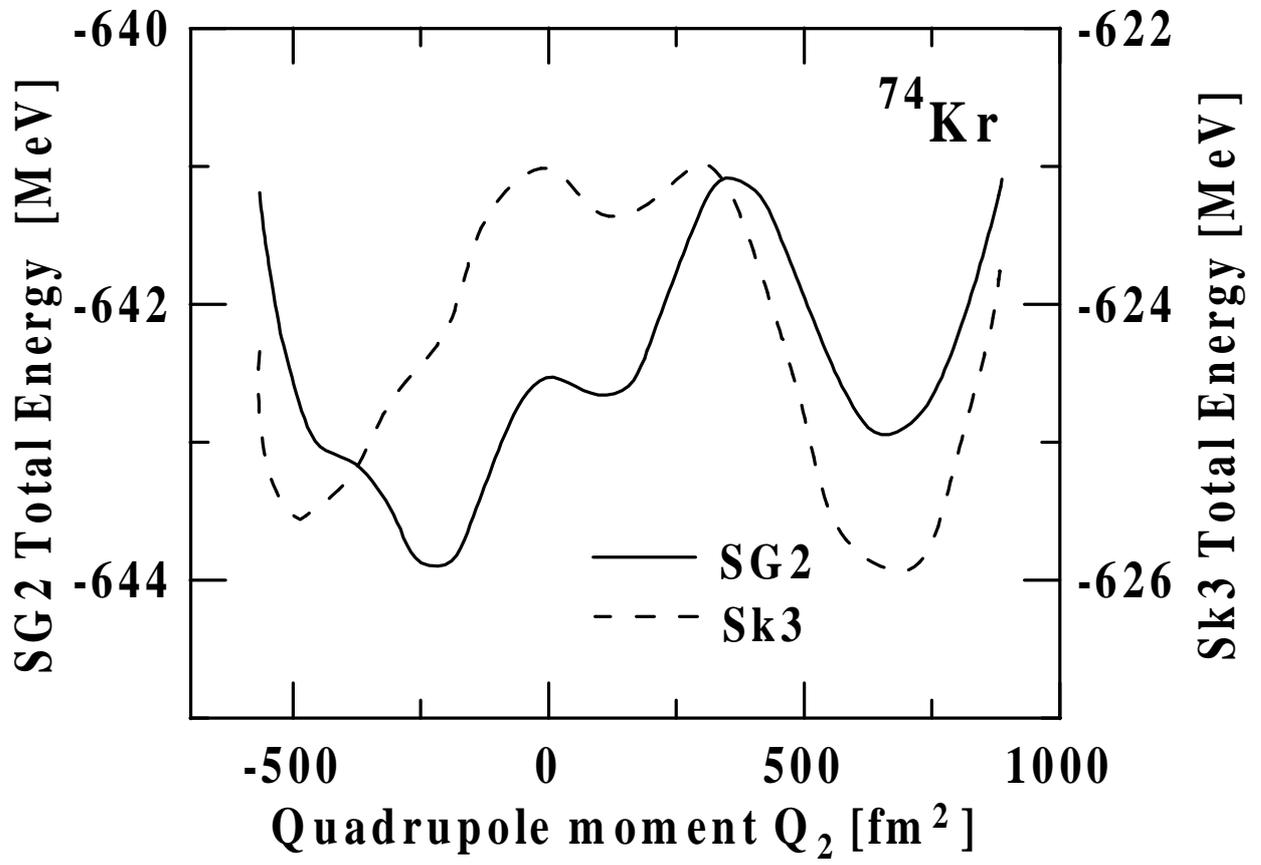
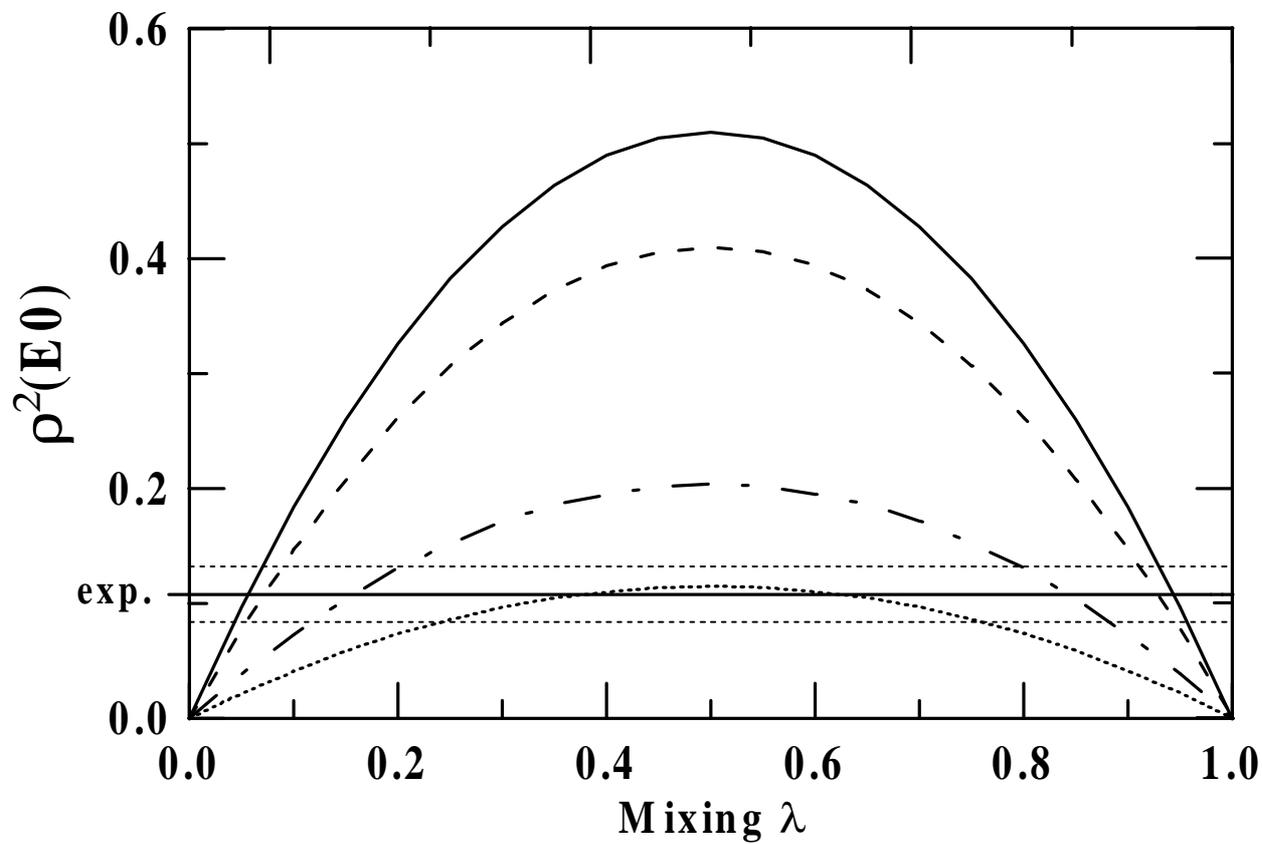
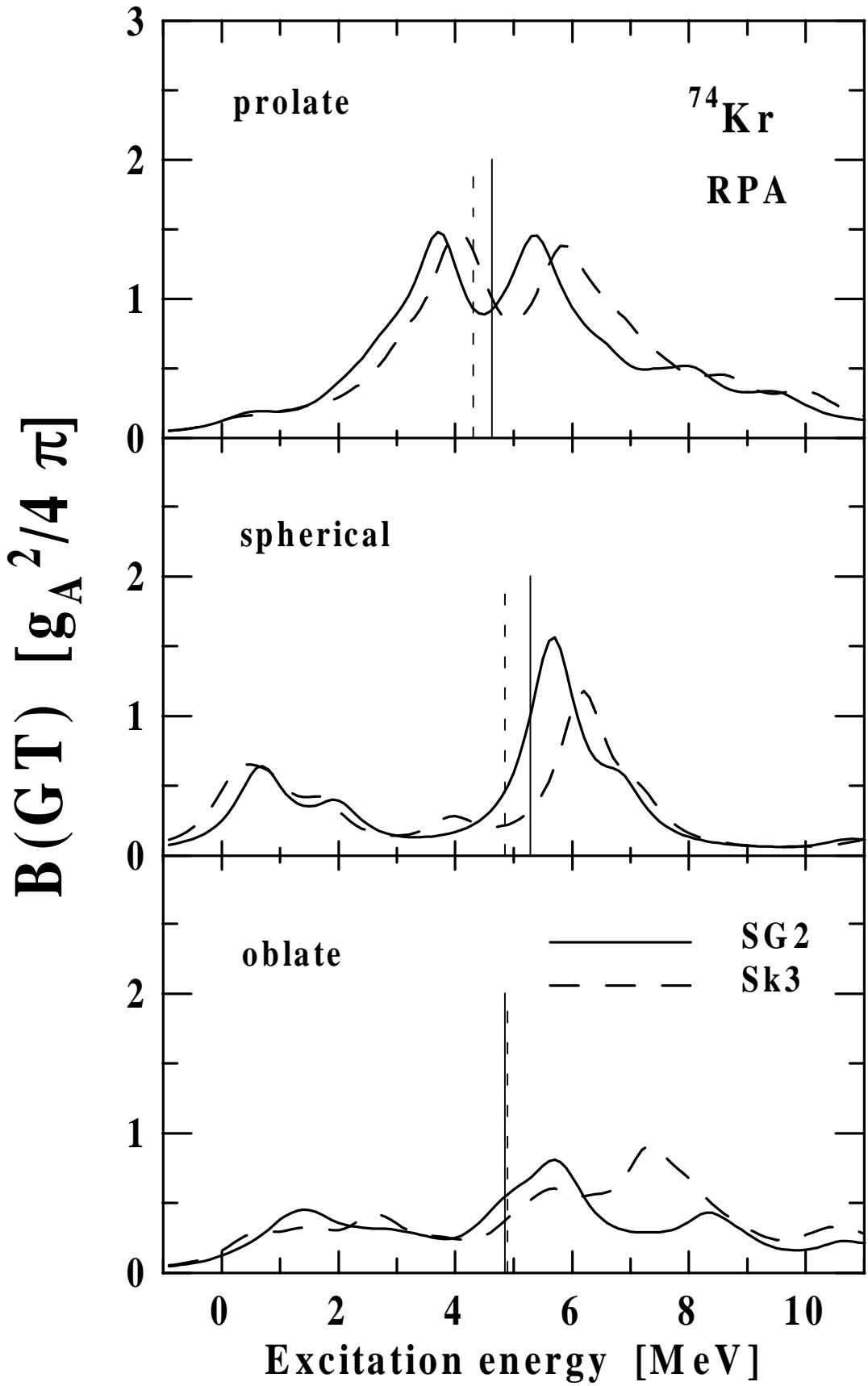
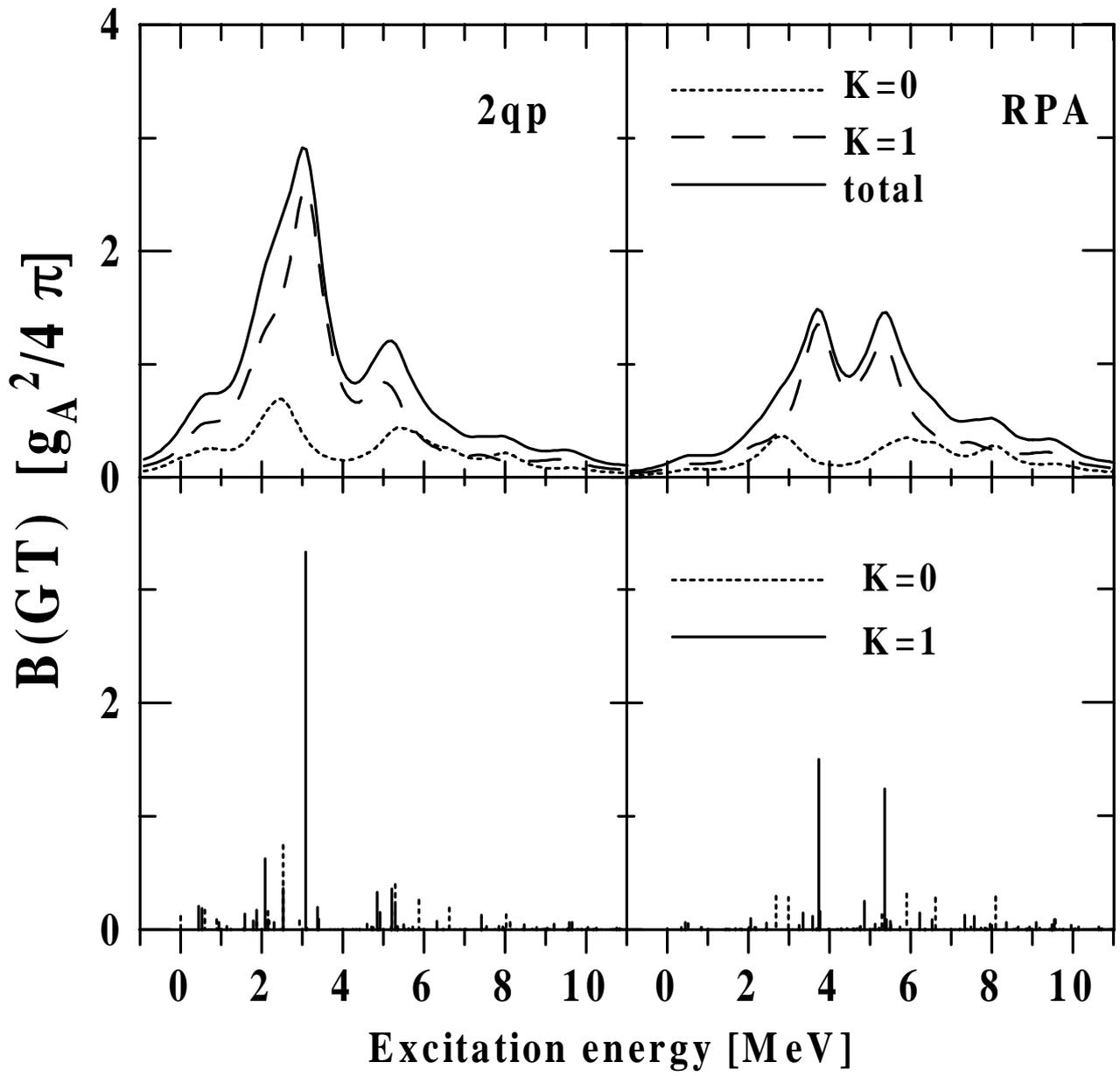
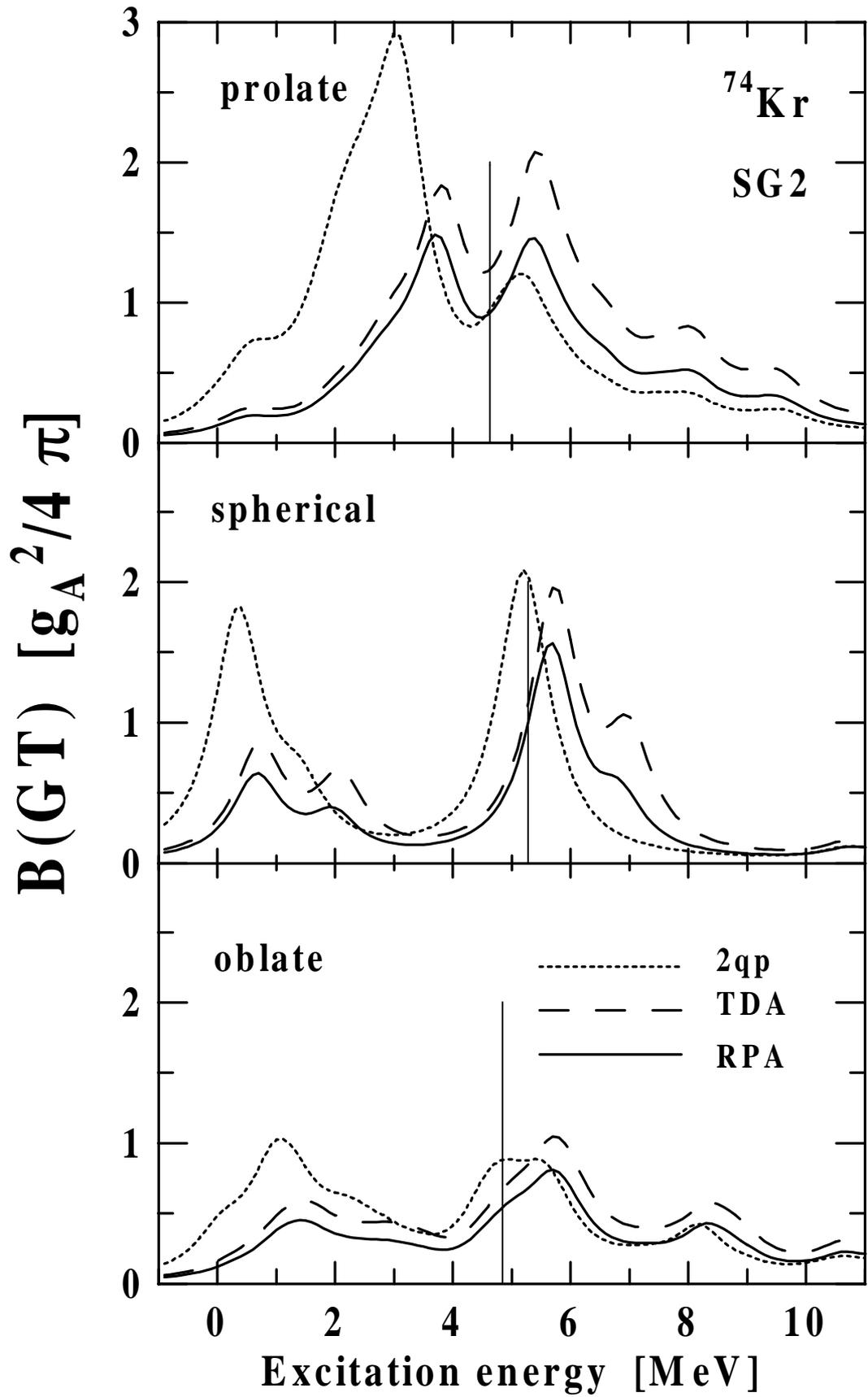
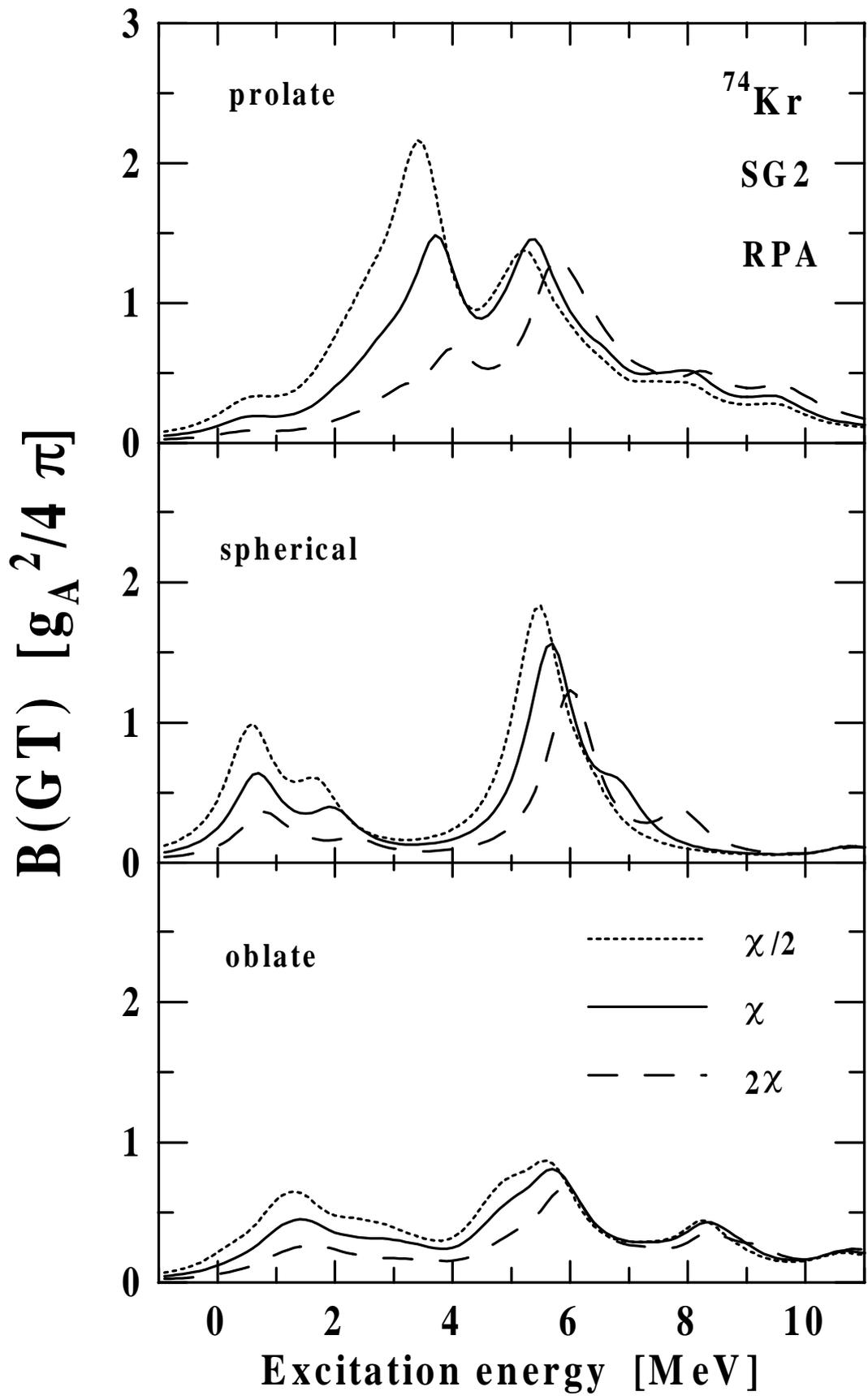
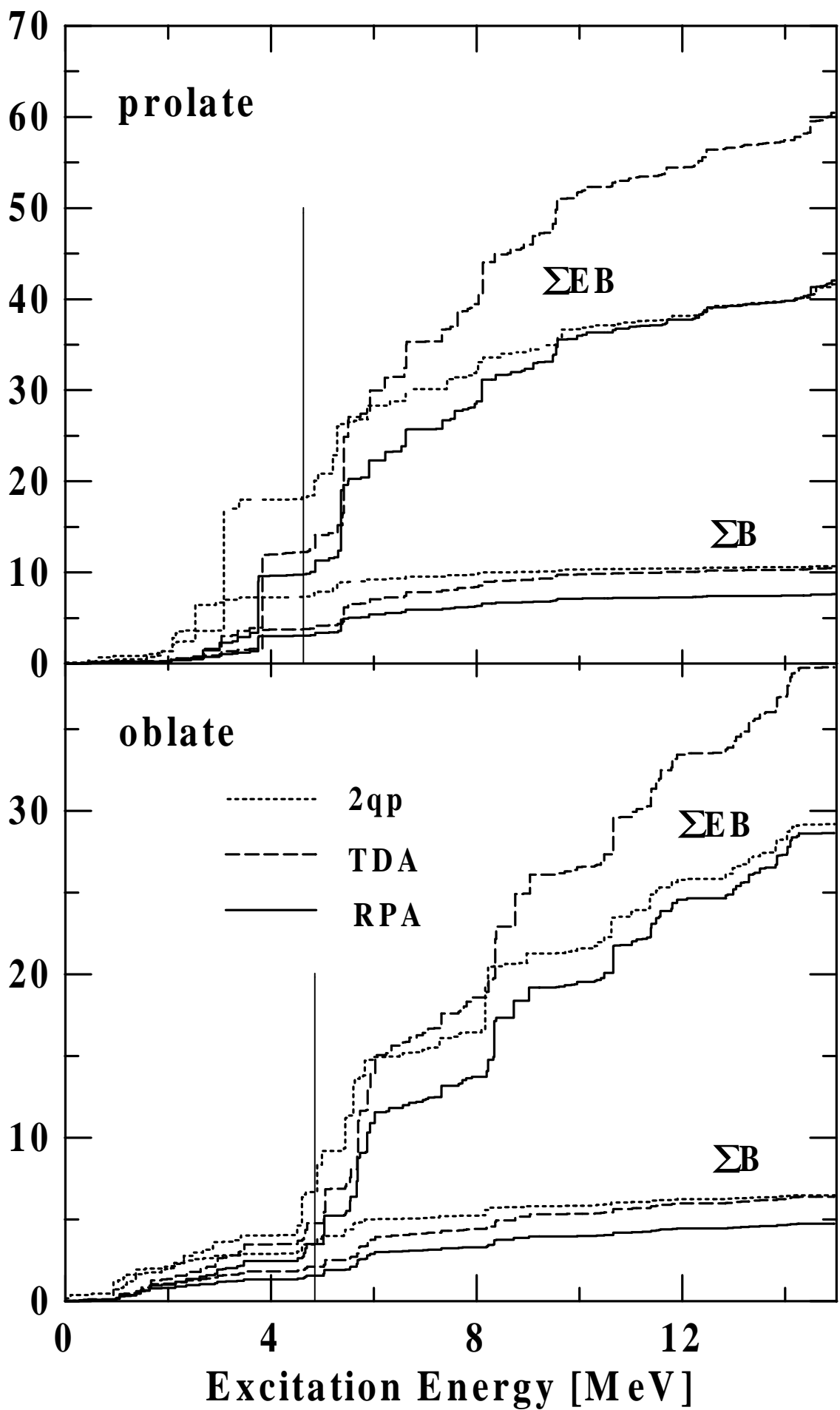
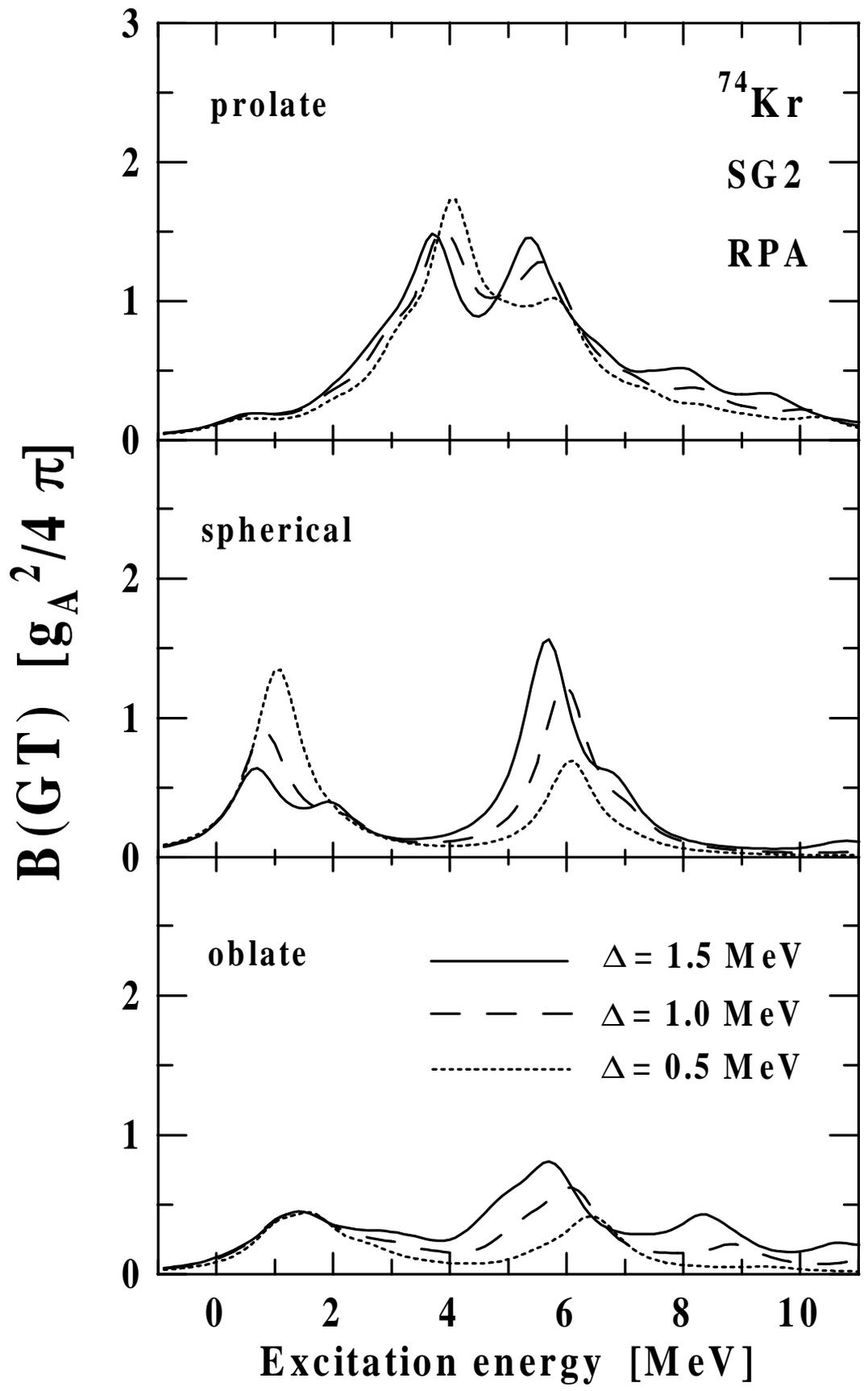
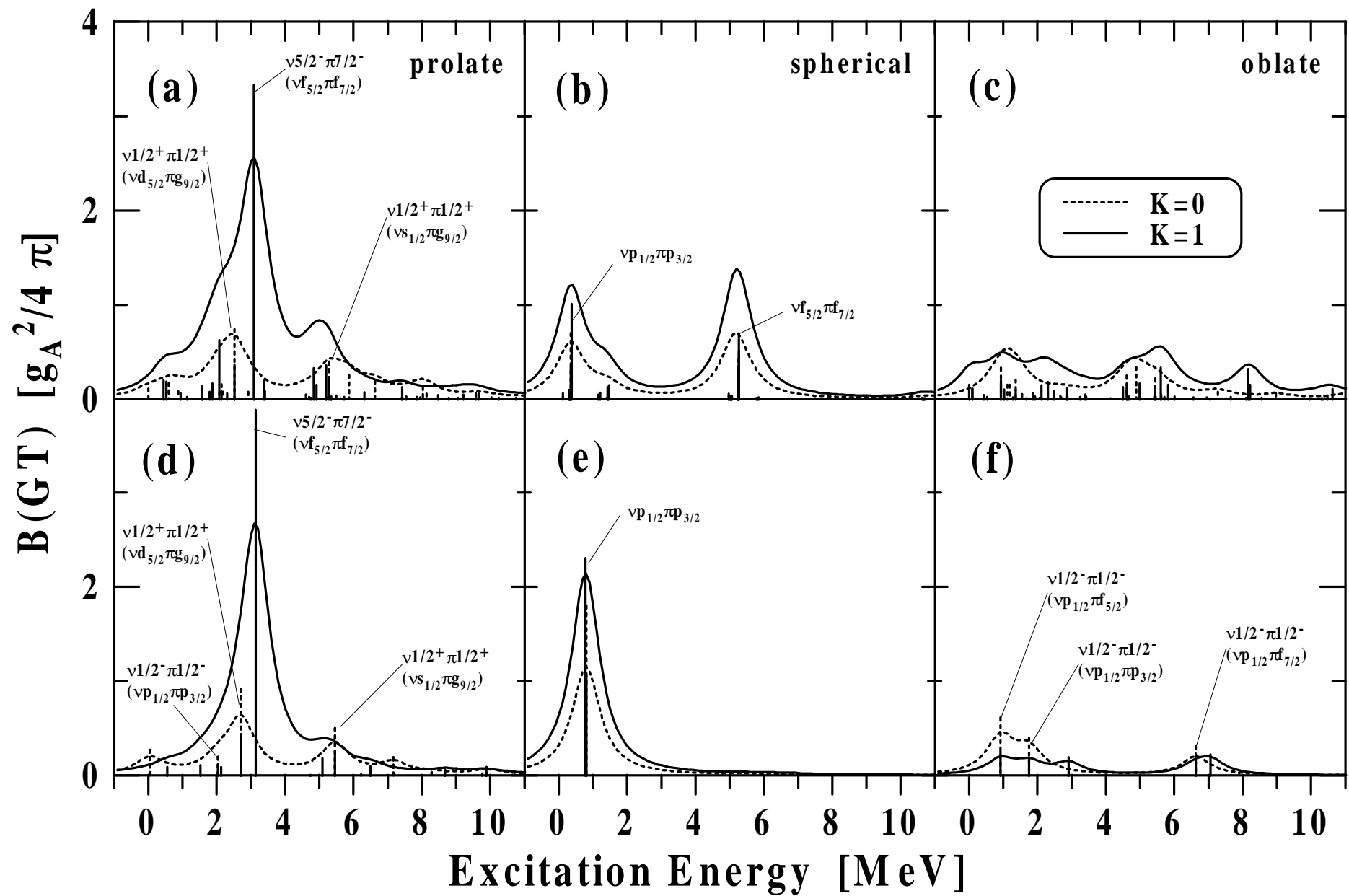
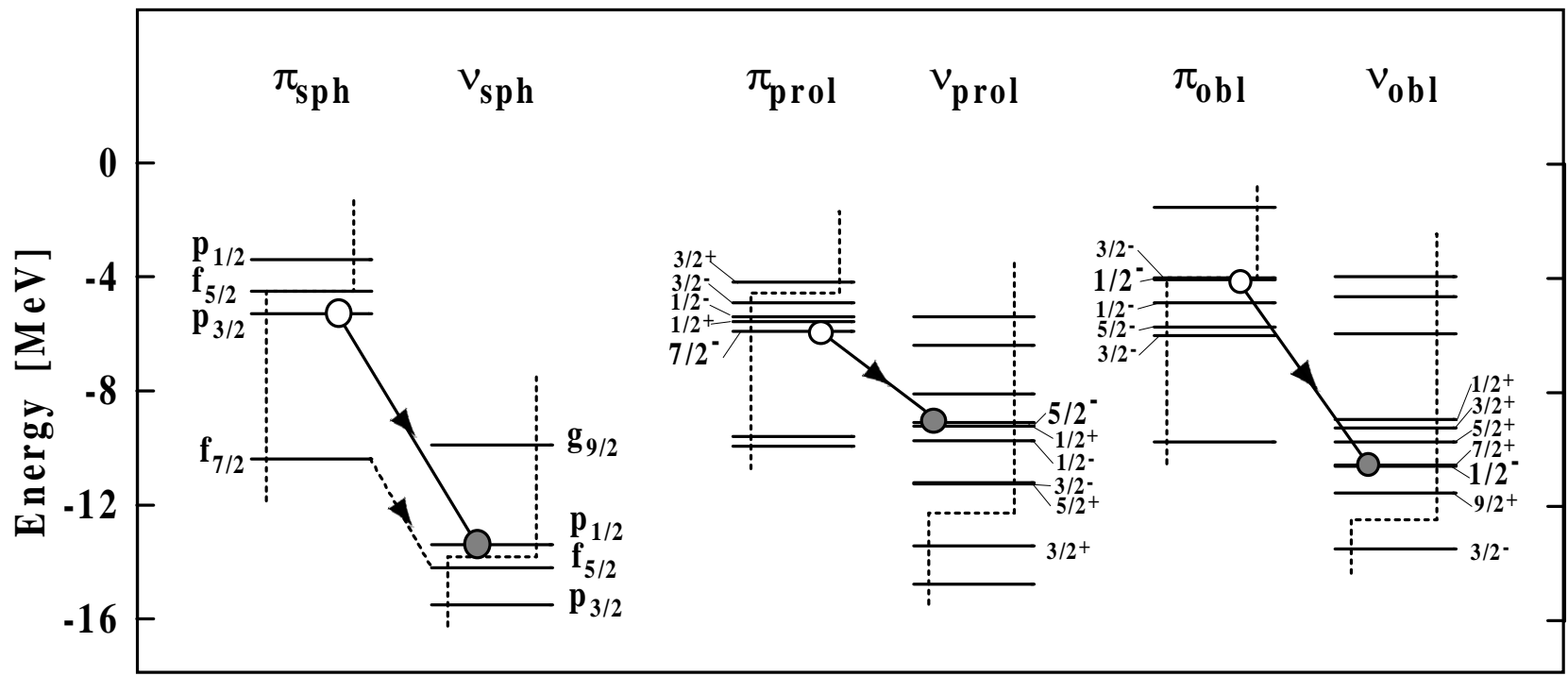
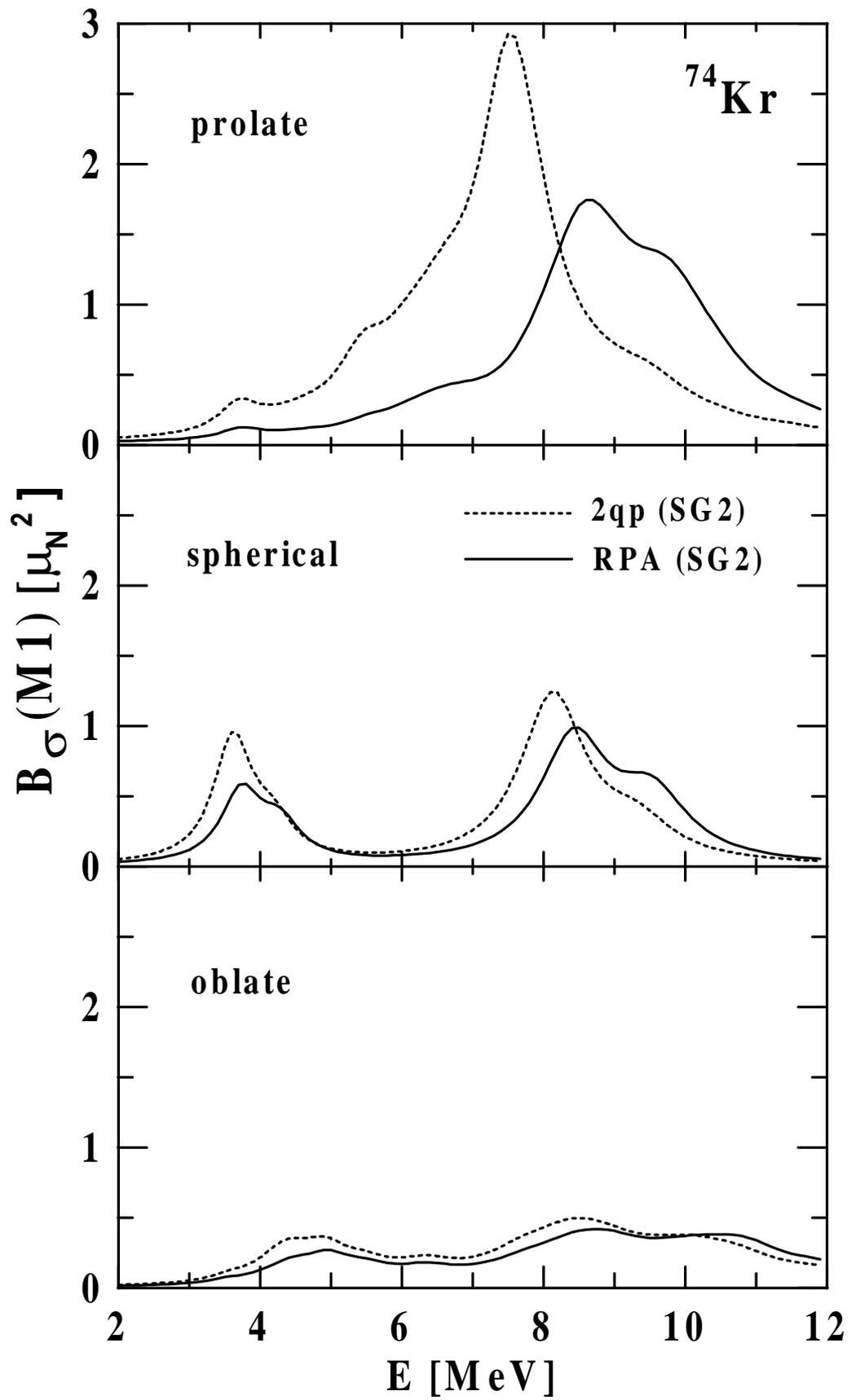
\end{document}